\documentclass[a4paper,fleqn,usenatbib]{mnras}
\usepackage{newtxtext,newtxmath}
\usepackage[T1]{fontenc}
\usepackage{ae,aecompl}
\usepackage{soul}

\usepackage{graphicx}   
\usepackage{amsmath}    
\usepackage{amssymb}    

\newcommand{\mstar}{$M_{\star}$} 
\newcommand{\msun}{M$_\odot$} 
\newcommand{\sag}{\textsc{sag}} 
\newcommand{\sagcal}{\textsc{sag$_{\beta1.9}$}} 
\newcommand{\oldfb}{\textsc{sag$_{\beta0.0}$}} 
\newcommand{\sagbeta}{\textsc{sag$_{\beta1.3}$}} 
\newcommand{\sagrec}{\textsc{sag$_{{\beta1.9-}{\rm Rec}}$}} 

\title[Semi-analytic galaxies III]{Semi-analytic galaxies -- III. 
The impact of supernova feedback on the mass-metallicity relation}
\author[F. Collacchioni et al.]{
Florencia Collacchioni$^{1,2,3}$\thanks{E-mail:fcollacchioni@fcaglp.unlp.edu.ar}, 
Sof\'ia A. Cora$^{1,2,3}$,
Claudia D. P. Lagos$^{4,5,6}$
and \newauthor{ Cristian A. Vega-Mart\'inez$^{1,3}$}
\\
$^{1}$Instituto de Astrof\'isica de La Plata (CCT La Plata, CONICET, UNLP), Observatorio Astron\'omico, Paseo del Bosque,\\ B1900FWA, La Plata, Argentina.\\
$^{2}$Facultad de Ciencias Astron\'omicas y Geof\'{\i}sicas, Universidad Nacional de La Plata, Observatorio Astron\'omico,\\ Paseo del Bosque, B1900FWA La Plata, Argentina.\\
$^{3}$Consejo Nacional de Investigaciones Cient\'ificas y T\'ecnicas (CONICET), Rivadavia 1917, Buenos Aires, Argentina.\\
$^{4}$International Centre for Radio Astronomy Research (ICRAR), M468, University of Western Australia, 35 Stirling Hwy,\\ Crawley, WA 6009, Australia.\\
$^{5}$Australian Research Council Centre of Excellence for All Sky Astrophysics in 3 Dimensions (ASTRO 3D).\\
$^{6}$Cosmic Dawn Center (DAWN), Niels Bohr Institute, University of Copenhagen$/$DTU-Space, Technical University of Denmark,\\ N{\o}rregade 10, 1165 K{\o}benhavn, Denmark.\\
}
\date{Accepted XXX. Received YYY; in original form ZZZ}
\pubyear{2018}
\begin{document}
\label{firstpage}
\pagerange{\pageref{firstpage}--\pageref{lastpage}}
\maketitle

\begin{abstract}
We use the semi-analytic model (SAM) of galaxy formation and evolution \sag~coupled with the \textsc{multidark} simulation MDPL2
to study the evolution of the stellar mass-gas metallicity relation of galaxies (MZR).
We test several implementations of the dependence of the mass loading due to supernovae (SNe). 
We find that no evolution in the normalization of the MZR is obtained unless we introduce an explicit scaling of the reheated 
and ejected mass with redshift as $(1+z)^\beta$.
The latter is in agreement with results from the FIRE simulations, and it should encompass small scale properties of the interstellar medium varying over time, which are not captured in SAMs, as well as other energy sources in addition to SNe. 
Increasing $\beta$ leads to stronger evolution of the MZR normalization; $\beta = 1.9$ reproduces the observed MZR in the range $0 < z < 3.5$.
A stronger redshift dependence of outflows reduces the levels of star formation at earlier epochs with the consequent decrease of metal production. This leads to a slower increase of the gas metallicity compared to the stellar mass build-up. The cold gas can be contaminated either by receiving a direct injection of the material recycled by stellar winds and SNe or by gas cooling. 
The relative role of each process for a given stellar mass depends on the criterion adopted to regulate the fate of the recycled material. 
However, modifying the metal loading of the outflows has mild impact on the zero-point evolution and does not affect our conclusions.
\end{abstract}



\section{Introduction}
\label{sec:Intro}
%
The gas-phase oxygen abundance, or gas metallicity, is the measure of the amount of oxygen relative to hydrogen in the gas content of a galaxy. 
Observations show that metallicity correlates positively with stellar mass \citep{Tremonti2004}. 
This relation is known as the mass-metallicity relation (MZR) and provides valuable insight onto the impact of the different physical processes that shape the formation and evolution of galaxies. 
Dying stars produce metals that are returned to the interstellar medium (ISM) via stellar mass loss, resulting in enrichment \citep{Zahid2012}, while gas dilution happens if inflows of pristine gas are present. 
Thus, these competing mechanisms leave imprints on the oxygen abundance that can be used to infer gas inflows \citep{Zahid2014b,Lagos2016}. 
Another important physical process that can affect the metallicity are galactic outflows, which can transport part of their metal to the galaxy halo and beyond \citep{AnglesAlcazar2017}.\\
\indent
A distinctive characteristic of the MZR is that at stellar masses $\gtrsim 10^{10}$ \msun~the metallicity flattens to an almost constant value \citep{Tremonti2004}. 
This effect has been attributed to galaxy downsizing \citep{Maiolino2008}, i.e., galaxies with higher stellar masses having their star formation rate (SFR) peaking at higher redshift, in which case the pollution rate of the cold gas in massive galaxies has dropped considerably compared to low mass galaxies, which are still forming stars and injecting new metals onto the ISM. 
\citet{Zahid2013} argue that the flattening can also be caused by metallicities saturating at an oxygen abundance equivalent to the yield produced by the stellar nucleosynthesis of stars that have enough time to evolve and return all those metals to the ISM. 
They go even further claiming that the saturation is produced by the equilibrium between the returned metals of dying stars (usually massive ones) and the metals still contained in living stars \citep{Zahid2014a}.

Many authors have found that the MZR does evolve in normalization, in a way that high redshift galaxies have a lower oxygen abundance at fixed stellar mass \citep{Erb2006, Maiolino2008, Nagao2008, Mannucci2009, Moller2013, Zahid2014a, Troncoso2014, Wu2016, Ly2016}. 
It is thought that the evolution of the MZR is a consequence of a complex interaction between the infall of pristine gas towards the galaxies, outflows of metal-rich gas, and mass loss from stars, as it has been suggested from the observations of atomic gas and its relation with metallicity \citep{Bothwell2013,Brown2018}.

Another interesting feature of the MZR is the observed scatter, even at $z=0$, as galaxies of the same stellar mass have a wide range of metallicities ($1 \sigma$ of  $\pm 0.3$~dex), according to the estimate done by \citet{Mannucci2010}. 
These authors found that introducing the SFR as an additional variable led to a significant reduction of the scatter to $0.05$~dex, with higher metallicities being associated to lower SFRs at fixed stellar mass. 
They were guided by the results obtained by \citet{Ellison2007} which show that the MZR has a mild negative correlation  with the specific SFR (sSFR) for galaxies with stellar mass $M_\star \lesssim 10^{10}$~\msun. 
At the same time, \citet{LaraLopez2010} introduce the concept of a Fundamental Plane that correlates the stellar mass, the gas-phase metallicity and the SFR, which seems not to evolve with time. 
This surface, called Fundamental Metallicity Relation (FMR) by \citet{Mannucci2010}, is of important interest because it could explain the dispersion of the MZR and provide insight into the study of galaxies at high redshift due to its invariance with time. 
However, the question of the role of the SFR in the MZR is a contingent one.

\citet{Sanchez2013} did not find a correlation between metallicity and SFR using integral field spectroscopy, arguing that the FMR was an artefact of the fibre spectroscopy used by previous work, which only sampled the central regions of galaxies, giving an incomplete view of their global metallicity. 
\citet{Brown2018} showed that depending on the adopted metallicity and SFR calibrations, the correlation between metallicity and SFR can go from being anti- to positively correlated.
Observations of atomic (HI) and molecular (H$_2$) hydrogen have been used to suggest that the gas content may be more fundamental than the SFR in determining the scatter of the MZR. 
On the same line, \citet{Zahid2014a} used an analytical model of chemical evolution to find that the relation between metallicity, stellar mass and hydrogen gas mass does not evolve with time for $z \lesssim 1.6$. 
Additionally, \citet{Zahid2014b} found observational evidence of the FMR evolution within the same redshift range. 
Hence, they conclude that the invariance of the FMR found by \citet{Mannucci2010} is attributed to the method used to calculate SFRs, and favour the idea that gas mass is a more fundamental property in the MZR. 
\citet{Bothwell2013,Hughes2013,Brown2018} have shown that higher metallicities are associated to lower gas fractions at fixed stellar mass. 
This naturally happens if gas plays the dual role of increasing the gas fraction as well as diluting the metals in the ISM as it inflows \citep{LaraLopez2013a}.

To reproduce the MZR locally and at high redshift has been proven to be a very challenging task for simulations. 
Semi-analytic models (SAMs) such as \textsc{galform} \citep{Cole2000, Bower2006, Gonzalez2014, Lagos2012, Lacey2016} and \textsc{l-galaxies} \citep{Springel2005, Croton2006, Guo2011, Henriques2015} do not produce a MZR that displays the observed normalization evolution \citep{Guo2016}.
An explanation for this behaviour might be that the metallicity of galaxies in these SAMs grows at the same rate as the stellar mass, on average.
Nonetheless, using the semi-analytic model \textsc{gaea} (\textit{GAlaxy Evolution and Assembly}), \citet{Xie2017} confirmed an evolution of the MZR up to $z \sim 0.7$ for all stellar masses and up to $z \sim 2$ for most massive galaxies (\mstar~$\gtrsim 10^{10}$ M$_\odot$), something that so far could not be achieved by SAMs, which they attributed to a combination of their outflow model and adopted star formation law.

Recent cosmological hydrodynamical simulations have had more success reproducing the MZR relation and its evolution. 
\citet{Dave2011} showed that changing the model of stellar winds in their Smooth Particle Hydrodynamic (SPH) code leads to different rates of MZR evolution due to the impact this has on the interplay between gas inflows and outflows. 
More recently, \citet{Dave2017} show that their new cosmological hydrodynamic simulation MUFASA displays an MZR that evolves in normalization by $0.4$~dex from $z=0$ to $z=6$.
Other cosmological hydrodynamical simulations, such as EAGLE (Evolution and Assembly of GaLaxies and their Environments; \citealt{Crain2015, Schaye2015}) show a similar magnitude of evolution \citep{Guo2016,DeRossi2017}. 
In addition, \citet{Lagos2016} and \citet{DeRossi2017} showed in EAGLE that the gas metallicity is more fundamentally correlated with gas fraction than SFR. 
Cosmological zoom-in simulations FIRE (Feedback in Realistic Environments; \citealt{Hopkins2013}) have also been able to reproduce the evolution of the MZR normalization \citep{Ma2016}, while discrepancies with observations become more prevalent in the slope of the MZR.

From these results it becomes apparent that hydrodynamical simulations are able to delay the metal enrichment of the ISM of galaxies to a degree that SAMs have struggled with. 
In this work we tackle this problem and investigate the origin of the evolution of the MZR with the semi-analytic model of galaxy formation and evolution \sag~\citep[][Paper I, hereafter]{Cora2018}, focusing on the impact produced by different modelling of supernova (SN) feedback, and how that changes the interplay between gas inflows/outflows.
\sag~features a detailed non-instantaneous recycling chemical enrichment model in which several elements, including oxygen, are followed individually. 
This makes \sag~an ideal laboratory to perform this study.

This paper is organized as follows.
In Sec.~\ref{sec:Model}, we give a brief overview of the dark matter (DM) simulation we use to couple with the semi-analytic model \sag~and of the physics included in this model (Paper I), describing with more detail the SN feedback model considered, and how the chemical enrichment of the different baryonic components is calculated.
We present the predicted MZR at $z=0$ and at higher redshifts, and compare them with observations in Sec.~\ref{sec:MZR}, showing that the model gives rise to evolution in the normalization of the MZR\footnote{Any time we mention MZR evolution we are referring to the evolution of the normalization of this relation.}.
In Sec.~\ref{sec:Analysis}, we analyse the results of two additional versions of the \sag~model that differ in the treatment of the reheated material by SNe with the aim of identifying the physical driver of the MZR evolution found in our reference model.
We also demonstrate that the explanation for the origin of the MZR zero-point evolution remains valid when changing the fate of the recycled material. 
We discuss our results and present our conclusions in Sec.~\ref{sec:Conclusions}.

\section{Galaxy formation model}
\label{sec:Model}
Our galaxy formation model combines the semi-analytic model of galaxy formation and evolution \sag~with a cosmological numerical simulation.
Semi-analytic galaxies are generated from merger trees built for all self-bound DM structures.
We use the \textsc{MultiDark} simulation MDPL2\footnote{\url{https://www.cosmosim.org}} \citep{Klypin2016}.
Properties of the  population of `MultiDark Galaxies' generated by \sag~are compared with those of other SAMs in \citet{Knebe2017a}. 
These galaxy catalogues are publicly available in the \textsc{CosmoSim} database\footnote{\url{http://dx.doi.org/10.17876/cosmosim/mdpl2/007}}.

The MDPL2 simulation follows the evolution of $3840^3$ DM particles with mass $m_\text{p}$~=~1.5~$\times$~10$^{9}\, h^{-1}\,$ \msun~ within a box of side-length $1\,h^{-1}\,{\rm Gpc}$, consistent with a flat $\Lambda$CDM model characterized by the Planck cosmological parameters: matter density $\Omega_{\rm m}$~=~0.307, energy density $\Omega_\Lambda$~=~0.693, baryon density $\Omega_{\rm B}$~=~0.048, spectral index $n_{\rm s}$~=~0.96 and Hubble constant $H_0$~=~100~$h^{-1}$~km~s$^{-1}$~Mpc$^{-1}$, with $h$~=~0.678 \citep{Planck2013}.
DM haloes have been identified with the \textsc{Rockstar} halo finder \citep{Behroozi2013a}, and merger trees were constructed with \textsc{ConsistentTrees} \citep{Behroozi2013b}.
Main host haloes are those detected over the background density and may contain subhaloes. 
Thus, each system of haloes hosts central and satellites galaxies as assigned by \sag. 
Orphan satellites are those galaxies that populate DM subhaloes that are no longer identified by the halo finder, which usually happens when satellite subhaloes fall deep in the potential well of a larger halo \citep{Elahi2018}.

Our model \sag~is based on the version presented by \citet{Springel2001}, and was modified as described in \citet{Cora2006}, \citet{Lagos2008}, \citet{Tecce2010}, \citet{Orsi2014}, \citet{MunnozArancibia2015} and \citet{Gargiulo2015}.
The latest improvements implemented are detailed in Paper I.
The main physical processes included in \sag~are the radiative cooling of hot gas within both main host haloes and subhaloes (strangulation is replaced by gradual starvation), star formation, feedback from SN explosions, growth of central supermassive black holes (BHs) and the consequent feedback from active galactic nuclei (AGN), and starbursts triggered by disc instabilities and/or galaxy mergers. 
Starbursts contribute to the bulge formation with the cold gas being consumed gradually as it is transferred to the bulge. 
Stars formed in both quiescent and bursty modes give place to stellar winds and SNe Type Ia and II that contribute with different chemical elements that contaminate all baryonic components through recycling, reheating, ejection and reincorporation processes; lifetime of SNe progenitors are taken into account. 
Gradual starvation on satellite galaxies is modelled through the action of ram pressure stripping (RPS) and tidal stripping (TS) on the hot gas halo. 
This gas phase can also be reduced by gas cooling and ejection, while it can be replenished by cold gas reheated by SN feedback and the reincorporation of ejected gas.
RPS and TS can also affect the cold gas disc when it is not longer shielded by the hot halo. 
Stellar mass from discs and bulges can also be removed by TS. 
Values of ram pressure are consistent with the position and velocity
of satellite galaxies since we consider a fitting formulae to estimate ram pressure as a function of halo mass, halo-centric distance and redshift. 
This improvement is complemented by the integration of the orbits of orphan galaxies according to the potential well of the host halo, taking into account mass loss by TS and dynamical friction effects (Vega-Mart\'inez et al., in preparation). 
Thus, our implementation of environmental effects presents two important developments that are missing in previous SAMs \citep{Gonzalez2014,Lacey2016, Henriques17, Stevens17}.

Since we are interested in studying how the metal enrichment of the cold gas varies depending on the details of SN feedback, we summarize the two variants of the modelling of this process in Section~\ref{sec:improve} (see Paper I for more details). 
We explain the chemical model implemented in \sag~in Section~\ref{sec:regulation}. 
Additionally, in Section~\ref{sec:calibration} we describe the calibration of our model.

\subsection{Modelling of SN feedback}
\label{sec:improve}
SN feedback is a crucial process that regulates the SFR, reheating the cold gas from which stars are formed.
In previous versions of~\sag, we assumed that the amount of reheated mass produced by the SNe arising in each star-forming event is
\begin{eqnarray}
        \Delta M_{\rm reheated} = \frac{4}{3} \epsilon {\frac{\eta E_\text{SN}}{V_{\rm vir}^2}} \Delta M_{\star},
        \label{eq:feedbackSN}
\end{eqnarray}
\noindent where $\eta$ is the number of SNe generated from the stellar population of mass $\Delta$\mstar~formed. 
$E_\text{SN}=10^{51}\,{\rm erg}$ is the energy released by a SNe, $V_{\rm vir}$ is the virial velocity of the host (sub)halo and $\epsilon$ is the SN feedback efficiency (a free parameter of the model). 
The quantity $\Delta M_{\rm reheated}$ is estimated for each time-step of equal size in which the time intervals between simulations outputs are subdivided to integrate differential equations.

This scheme of SN feedback is modified in the current version of \sag~with the purpose of satisfying observational constraints at high redshift (Paper I) by adding new factors that take into account an explicit dependence on redshift and an additional modulation with virial velocity, as suggested by the cosmological hydrodynamical zoom-in FIRE simulations \citep{Muratov2015}. 
Therefore, the feedback scheme produces a reheated gas mass given by
\begin{eqnarray}
        \Delta M_{\rm reheated} = \frac{4}{3} \epsilon {\frac{\eta E_\text{SN}}{V_{\rm vir}^2}}\,(1+z)^{\beta}\,\left(\frac{V_{\rm vir}}{60\,{\rm km}\,{\rm s}^{-1}}\right)^{\alpha}\Delta M_{\star},
        \label{eq:feedfire}
\end{eqnarray}
\noindent where the exponent $\alpha$ takes the values $-3.2$ and $-1.0$ for virial velocities smaller and larger than $60\,{\rm km}\,{\rm s}^{-1}$, respectively.
The efficiency, $\epsilon$, and power-law slope of the redshift dependence, $\beta$, are free parameters of \sag. 

The reheated gas is transferred from the cold to the hot phase, i.e., part of the mass in the cold gas disc, which represents the ISM in our model galaxies, is heated and relocated in the hot gas halo of the galaxy. 
Part of it returns afterwards to the cold phase through gas cooling, a process that takes place in both central and satellite galaxies.
We found that some hot gas must be ejected out of the halo reducing the hot gas reservoir available for gas cooling in order to avoid an excess of stellar mass at high redshifts \citep{Guo2011, Henriques2013, Hirschmann2016}.
Therefore, the new SN feedback model is complemented with the ejection of hot gas mass. 
The ejected mass is calculated as
\begin{eqnarray}
        \Delta M_{\rm ejected}= \frac{\Delta E_\text{SN} - 0.5\,\Delta M_{\rm reheated}\,V_{\rm vir}^2}{0.5\,V_{\rm vir}^2},
        \label{eq:EjecMass}
\end{eqnarray}
\noindent where $\Delta E_\text{SN}$ is the energy injected by massive stars \citep{Guo2011}. 
We model this energy in a way similar to the modified reheated mass
\begin{eqnarray}
        \Delta E_{\rm SN} = \frac{4}{3} \epsilon_\text{ejec} {\frac{\eta E_\text{SN}}{V_{\rm vir}^2}} \,(1+z)^{\beta}\,\left(\frac{V_{\rm vir}}{60\,{\rm km}\,{\rm s}^{-1}}\right)^{\alpha}\Delta M_{\star}\,0.5\,V_\text{SN}^2,
        \label{eq:energySN}
\end{eqnarray}
\noindent where $\epsilon_\text{ejec}$ is the corresponding efficiency, considered as a free parameter of the model, and $0.5\,V_\text{SN}^2$ is the mean kinetic energy of SN ejecta per unit mass of stars formed, given by $V_\text{SN}=1.9\,V_\text{vir}^{1.1}$ \citep{Muratov2015}.

With the aim of reproducing the observed evolution of the galaxy stellar mass function at $z \geqslant 1$, the (sub)halo needs to re-incorporate the previously expelled ejected gas mass within a timescale that depends on the inverse of the (sub)halo mass, $M_{\rm vir}$ \citep{Henriques2013}. 
In this way, the reincorporated mass is given by
\begin{eqnarray}
        \Delta M_{\rm reinc}= \gamma\,\Delta M_\text{ejected}\,\frac{10^{10}\,{\rm M_{\odot}}}{M_{\rm vir}},
        \label{eq:reinc}
\end{eqnarray}
\noindent where $\gamma$ is a free parameter that regulates the efficiency of the process.

\subsection{
Exchange of mass and metals between different baryonic components
}
\label{sec:regulation}
The way in which the exchange of mass and metals between the different baryonic components takes place is a key aspect in the interpretation of the results on the MZR evolution. 
The chemical model implemented in \sag~was originally introduced by \citet{Cora2006}. 
Here, we provide a concise description adapted to the updated version of the model since we now consider ejection and reincorporation of the hot gas phase.

We assume that the hot gas always has a distribution that parallels that of the main host dark matter halo, whose virial mass, $M_{\rm vir}$, changes between consecutive outputs of the underlying simulation as a result of the hierarchical growth of the structure. 
Initially, the mass of hot gas is given by $M_{\rm hot}=f_{\rm b} M_\text{vir}$, where $f_{\rm b}$~=~0.1569 is the universal baryon fraction \citep{Planck2013}. 
With this definition, we model the cosmological smooth accretion that takes place onto main host haloes as their virial masses increase. 
The accreted gas is assumed to have a primordial chemical composition ($76$ per cent of hydrogen and $24$ per cent of helium). 
Once a fraction of hot gas has cooled and star formation and feedback processes are triggered, the mass of hot gas available for cooling to the central galaxy of each main host halo is calculated as
\begin{equation}
  \begin{split}
    M_\text{hot} = & f_{\rm b} M_\text{vir} - M_{\star,\text{cen}} -
    M_\text{cold,cen} - M_\text{BH,cen} \\
    & - \sum_{i=1}^{N_\text{sat}} \left( M_{\star,i} +
    M_{\text{cold,}i} + M_{\text{BH,}i} + M_{\text{hot,}i} \right),
  \end{split}
  \label{eq:hotgas}
\end{equation}
where $M_{\star}$, $M_{\rm cold}$ and $M_{\rm BH}$ are, respectively, the total masses of stars, cold gas and  central BHs of each of the galaxies contained within a given main host halo, either the central or the $N_\text{sat}$ satellites; for the latter, the total mass in the remaining hot gas haloes is also discounted.

For a given mass of stars formed, $\Delta M_\star$, the fraction of stellar mass within a certain mass range is calculated by assuming an initial mass function (IMF), $\Phi$(M).
\sag~tracks the mass losses of low, intermediate and high-mass stars that take place through stellar winds and SN explosions. 
In the latter case, the lifetime of SN progenitors is considered, thus relaxing the instantaneous recycling approximation. 
Stars contribute both recycled and newly synthesized material that gives place to the chemical enrichment of the gas phases and can result in an increase of the metallicity of future generations of stars. 
The mass of the chemical element $j$ ejected by stars with masses in the range centred in $m_{k}$ is given by
\begin{eqnarray}
\Delta M_{{\rm ej}_k}^j = [R_{k} X^j + Y_{k}^j] \Delta M_\star,
\label{eq:MassEjected}
\end{eqnarray}
where $X^j$ is the solar abundance of element $j$, $R_k$ is the recycled fraction, and $Y_k^j$ is the yield of element $j$; the two latter quantities depend on the $k$th mass range. The recycled fraction is given by
\begin{eqnarray}
R_{k} = \int_{m_{k}^l}^{m_{k}^u} \Phi(M) r_{k} {\rm d}M,
\label{eq:RecycledFrac}
\end{eqnarray}
where $m^l_k$ and $m^u_k$ depict lower and upper mass limits, and $r_k$ is the difference between the mass $m_k$ of the star at birth and its remnant mass after mass loss due to stellar winds and/or SN II explosions. 
The yield of the newly formed chemical element $j$ is calculated as
\begin{eqnarray}
Y_k^j = \int_{m_k^l}^{m_k^u} \Phi(M) p_k^j {\rm d}M,
\label{eq:YieldJ}
\end{eqnarray}
where $p_k$ represents the stellar yield of a chemical element $j$ produced by a star with mass within the range centred around $m_k$.

Due to the fact that SNe Ia do not leave remnants, the corresponding ejected mass is estimated by a different expression as
\begin{eqnarray}
\Delta M_{\rm ej(Ia)_k}^j = \eta_{{\rm Ia}_k} m_{{\rm Ia}_k}^j \Delta M_\star
\label{eq:EjectedIa}
\end{eqnarray}
where $m_{{\rm Ia}k}^j$ is the mass of species $j$ ejected by each SN Ia, and $\eta_{{\rm Ia}k}$ is the number of SN Ia per stellar mass (which depends on the model adopted for this type of SN), both depend on the mass interval centred in $m_k$.

The material ejected through stellar winds and SN explosions constitutes the recycled mass, $\Delta M_{\rm recycled}=\Delta M_{\rm ej} + \Delta M_{\rm ej(Ia)}$, which is the result of the accumulated contribution of all chemical elements ejected by stars in different mass ranges (equations~\ref{eq:MassEjected} and~\ref{eq:EjectedIa}) at a given time determined by the combination of the SFR of each galaxy and the return time-scale of all sources considered. 
The newly synthesized metals contained in $\Delta M_{\rm recycled}$ contribute towards enhancing the chemical abundance of all baryonic components through outflows and inflows.

The injected energy resulting from SN explosions reheats a fraction of the cold gas, referred to as reheated mass ($\Delta M_{\rm reheated}$ given by equations~\ref{eq:feedbackSN} or~\ref{eq:feedfire}, depending on the SN feedback model used).
This is transferred to the hot gas halo as an outflow of chemically enriched gas with the cold gas metallicity.
The new SN feedback scheme allows the ejection and later reincorporation of the hot gas ($\Delta M_{\rm ejected}$ and $\Delta M_{\rm reinc}$ given by equations~\ref{eq:EjecMass} and~\ref{eq:reinc}, respectively).
Metal-dependent cooling rates are taken into account to calculate the amount of cooled gas, $\Delta M_{\rm cool}$, whose metallicity is determined by the chemical abundance of the hot gas, and is transported from this phase to the cold gas disc, feeding the cold gas reservoir from which new stars are formed. 
In this scenario, the variation of the mass of  chemical element $j$ contained in a given baryonic component (hot gas, cold gas, stars and reservoir of ejected gas) is given by
\begin{eqnarray}
\Delta M_{\rm hot}^j = - \Delta M_{\rm cool} A_{\rm hot}^j + \Delta M_{\rm reheated} A_{\rm cold}^j 
\nonumber \\
+ f_{\rm rec,h} \left(\Delta M_{\rm ej}^j + \Delta M_{\rm ej(Ia)}^j \right)
\nonumber \\
- \Delta M_{\rm ejected} A_{\rm hot}^j +  \Delta M_{\rm reinc} A_{\rm ejected}^j,
\label{eq:MhotJ}
\end{eqnarray}
\begin{eqnarray}
\Delta M_{\rm cold}^j = + \Delta M_{\rm cool} A_{\rm hot}^j - \Delta M_\star A_{\rm cold(SF)}^j 
\nonumber \\ 
+ f_{\rm rec,c} \left(\Delta M_{\rm ej}^j + \Delta M_{\rm ej(Ia)}^j\right) - \Delta M_{\rm reheated} A_{\rm cold}^j,
\label{eq:McoldJ}
\end{eqnarray}
\begin{eqnarray}
\Delta M_{\rm stellar}^j = + \Delta M_\star A_{\rm cold(SF)}^j - \Delta M_{\rm recycled} A_{\rm star}^j,
\label{eq:MstarJ}
\end{eqnarray}
\begin{eqnarray}
\Delta M_{\rm ejected}^j = 
\Delta M_{\rm ejected} A_{\rm hot}^j -  \Delta M_{\rm reinc} A_{\rm ejected}^j,
\label{eq:MejecJ}
\end{eqnarray}
\noindent where $A_{\rm B}^j = M_{\rm B}^j/M_{\rm B}$ is the abundance of the chemical element $j$ in the baryonic component B, with $M_{\rm B}^j$ the mass of element $j$ contained in the component B of total mass $M_{\rm B}$. 
The suffix B alludes to cold gas, hot gas, stars and ejected gas. In the latter case, the abundance is a result of the succession of ejection events that combines different hot gas metallicities. 
The abundance $A^j_{\rm cold(SF)}$ represents the mass content of element $j$ in the cold gas at the time of the birth of stars with total stellar mass $\Delta M_\star$.
The fraction $f_{\rm rec,x}$ denotes the fraction of the recycled mass (combination of recycled and newly synthesized chemical elements) that is added into the hot (${\rm x=h}$) and cold (${\rm x=c}$) phases.

The fate of the recycled material, which contains metals that had been locked up in stars and newly synthesized chemical elements, is determined by the relative amounts of  reheated and recycled mass. 
The latter can either be kept within the cold gas disc or injected into the hot gas halo.
Both the reheated and recycled mass are a result of a single process, SN explosions.
We avoid the two-steps procedure in which the recycled material is first completely diluted in the cold gas and then a fraction of it is transferred to the hot phase because of the energy injection.
We consider that if the reheated mass is smaller than the recycled mass, then a fraction $f_{\rm rec,h}=\Delta M_{\rm reheated}/\Delta M_{\rm recycled}$ of the recycled material is added to the hot gas halo, and the rest ($f_{\rm rec,c}=1-f_{\rm rec,h}$) remains in the cold gas disc (see third term of equations~\ref{eq:MhotJ} and ~\ref{eq:McoldJ}).
On the contrary, if the reheated mass is higher than the recycled mass, all of the latter is directly added to the hot gas halo ($f_{\rm rec,h}=1$ and $f_{\rm rec,c}=0$), meaning that it has been completely reheated.\\
\indent
We also evaluate a variant of the \sag~model that considers a simpler criterion to decide the fate of the recycled mass. 
In this new scheme, the fraction of recycled material that is added to the cold gas phase is proportional to the cold gas fraction of the galaxy, i.e. the ratio between the cold gas content and the sum of cold gas and stars.
Thus, the fractions involved in the third term of equations~\ref{eq:MhotJ} and~\ref{eq:McoldJ} are replaced by $f_{\rm rec,c}=M_{\rm cold}/(M_{\rm cold}+M_{\star})$ and $f_{\rm rec,h}=1-f_{\rm rec,c}$.\\
\indent
In any of these two schemes, the chemical elements contained in the recycled mass are assumed to instantly mix with the metals of the corresponding gas phase.
Thus, the cold gas enhances its metal abundance through two channels, namely, the direct contribution of recycled material and the infall of enriched cooled gas, whose metallicity is determined by the metallicity of the hot gas.
At a certain time, all the baryonic components have a uniform distribution of mass and metals, since \sag~does not track the radial dependence of these quantities. 
The latter is a common feature of SAMs.\\
\indent
\sag~follows the abundance of eleven different chemical elements (H, $^4$He, $^{12}$C, $^{14}$N, $^{16}$O, $^{20}$Ne, $^{24}$Mg, $^{28}$Si, $^{32}$S, $^{16}$Ca, $^{56}$Fe). 
We consider the stellar mass dependent yields $p_k$ specified in \citet{Gargiulo2015}, estimated for a Chabrier IMF \citep{Chabrier2003}.
We adopted the total ejected mass of solar metallicity for all sets of stellar yields, assuming the solar abundances of \citet{Anders1989} with a solar composition of $Z_\odot = 0.02$.
Including the stellar metallicity dependence of the yields in the model has a very minimal effect on the galaxy properties studied here, while increasing the computational expense of the model significantly.
Return time-scales of material ejected by all sources considered are obtained from the stellar lifetimes given by \citet{Padovani1993}. 
Gas cooling rates are estimated considering the total radiated power per chemical element given by \citet{Foster2012}.\\
\indent
Unless otherwise stated, throughout the paper we refer to 
the oxygen-to-hydrogen abundance of the interstellar medium of galaxies in the model as gas metallicity. 
We calculate this metallicity as $12+\rm log(O/H)$, where $\rm O/H$ is the oxygen mass in the gas relative to that of hydrogen. 
This definition implicitly means we are calculating a gas mass-weighted oxygen abundance, while in observations this measurement is weighted by the luminosity of HII regions. 
This could introduce some systematic differences between our measurement and the observational one. 
However, since we are interested in the evolutionary trends of the MZR, we use observations only as a guidance and we do not attempt to correct our measurement to mimic the observational one.

\subsection{Calibration of \textsc{SAG} and model definitions}
\label{sec:calibration}
The efficiency of the physical processes implemented in the \sag~model is regulated by free parameters that are tuned by imposing observed galaxy properties as constraints.
The calibration of \sag~is performed through the Particle Swarm Optimization (PSO) technique \citep{Ruiz2015} adopting a set of five observational constraints defined in \citet{Knebe2017b}. 
We use the data compiled by \citet{Henriques2015} for the stellar mass functions (SMF) at $z=0$ and $z=2$, data presented by \citet{Gruppioni2015} for the SFR distribution function in redshift interval $z \in [0.0,0.3]$, data from \citet{Boselli2014} for the fraction of mass in cold gas as a function of stellar mass, and the combined datasets from \citet{McConnell2013} and \citet{Kormendy2013} for the relation between bulge mass and the mass of the central supermassive BH.

\begin{figure}
	\includegraphics[width=1\columnwidth]{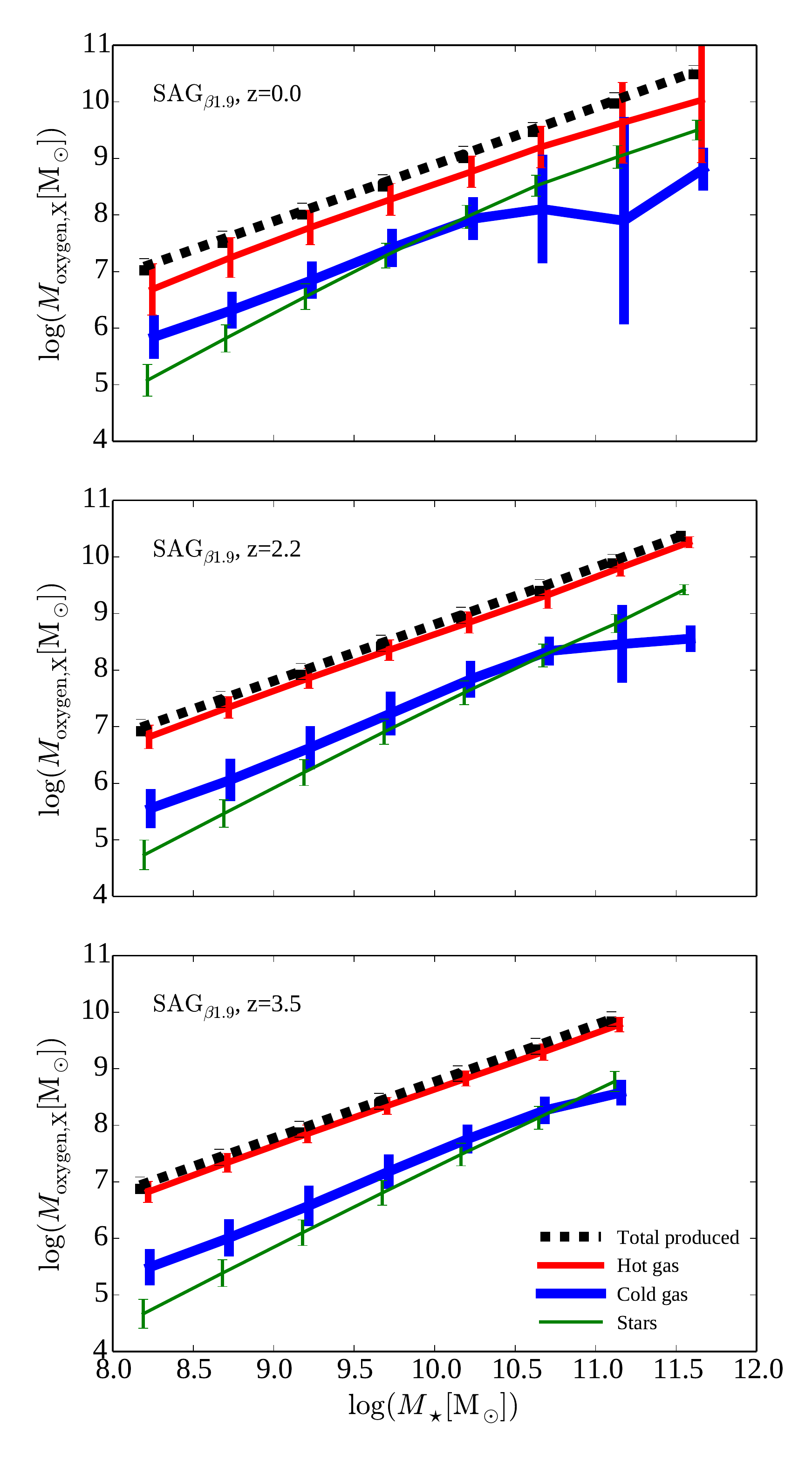}
    \caption{
    Oxygen mass of SF galaxies for \sagcal~model as a function of stellar mass at $z=0$ (top panel), $z=2.2$ (middle panel) and $z=3.5$ (bottom panel). 
    The black dashed, the red solid, the blue thickest solid and the green thinnest solid lines depict, respectively, the mean values of the total amount of oxygen produced by the galaxy through stellar nucleosynthesis, the oxygen mass in hot gas, the oxygen mass in cold gas, and the amount of oxygen locked up in stars. 
    The $1\,\sigma$ dispersion is represented by error-bars. 
    At all redshifts, the majority of the oxygen mass is in the hot gas phase.
    }
    \label{fig:OxygenMass}
\end{figure}

The best-fitting values of the free parameters of \sag~are shown in Table \ref{ta:param}. 
These parameters are the star formation efficiency ($\alpha$), the efficiency of SN feedback ($\epsilon$), those involved in the new feedback scheme (Eq.~\ref{eq:feedfire}, \ref{eq:energySN} and \ref{eq:reinc}), i.e., the efficiency of ejection of gas from the hot phase ($\epsilon_{\rm ejec}$) and of its reincorporation ($\gamma$), and the power-law slope of the redshift dependence ($\beta$), the growth of super massive BHs ($f_\text{BH}$) and efficiency of AGN feedback ($\kappa_\text{AGN}$), the factor that regulates the trigger of disc instabilities through the presence of a perturbing galaxy within a certain distance scale ($f_{\rm pert}$), and the fraction of hot gas content of satellite galaxies that determines the destination of the reheated cold gas ($f_\text{hot,sat}$), so that if the hot gas reservoir drops below this threshold, the reheated gas is transferred to the respective central galaxy.
The equations involving these parameters can be found in \citet{Ruiz2015} and Paper I, together with a description of the corresponding physical processes. 
Paper I also presents the details of the procedure followed to calibrate \sag~onto the MDPL2 simulation and the behaviour of statistics and relations followed by the properties of model galaxies. 
\begin{table}
  \centering
  \begin{tabular}{l r r}
    \hline
    Parameter                   & \sag \\
    \hline
    $\alpha$                    & 0.0402 \\
    $\epsilon$                  & 0.3299 \\
    $\epsilon_{\rm ejec}$       & 0.02238 \\
    $\gamma$                    & 0.0555 \\
    $\beta$             & 1.99  \\
    $f_{\rm BH}$                & 0.0605 \\
    $\kappa_{\rm AGN}$          & 3.02 $\times$ 10$^{-5}$ \\
    $f_{\rm pert}$              & 14.557 \\
    $f_{\rm hot,sat}$           & 0.2774 \\
        \hline
  \end{tabular}
  \caption{
  Best-fitting values of the free parameters obtained with the PSO technique for the \sag~model. 
  These values define the variant of the model to which we refer to as \sagcal. 
}
      \label{ta:param}
\end{table}
The calibration of \sag~gives $\beta=1.99$, a value higher than the one suggested by \citet[][$\beta=1.3$]{Muratov2015}. 
As discussed in Paper I, this value is a consequence of the constraint imposed by the low mass end of the SMF at $z=2$, which gives rise to an underprediction of the cosmic star formation rate density (SFRD) at high redshifts (see Sec.~\ref{sec:coldgas}). 
Model predictions, such as SFRD and the fraction of passive galaxies as a function of stellar mass, halo mass and halocentric distance, are in better agreement with observational data when adopting $\beta=1.3$, while keeping the best-fitting values of the rest of the free parameters obtained from the calibration process.
This variant of the model is referred to as the model \sagbeta~in Paper I.\\
\indent
The analysis of the calibrated model \sag~and the model \sagbeta~allows us to evaluate the impact of the different strength of SN feedback at high redshifts on the development of the MZR.
We will refer to the former as \sagcal, hereafter.
We also analyse the influence of the changes introduced in the modelling of the SN feedback itself by considering a version of \sag~in which the current feedback model (Eq.~\ref{eq:feedfire}) is replaced with the original prescription adopted for this process (Eq.~\ref{eq:feedbackSN}), suppressing ejection and reincorporation and keeping the same best-fitting values of the free parameters of the calibrated model \sag.
We refer to this run as~\oldfb. 
Note that this label not only means that the reheated mass loses the explicit redshift dependence but it also indicates the lack of the additional modulation with virial velocity of Eq.~\ref{eq:feedfire}.
In order to evaluate the impact of the scheme adopted to decide the fate of the recycled mass on the MZR evolution, we define a new variant of the model in which the corresponding scheme in \sagcal~is replaced by the simpler recipe that relies on the cold gas fraction; this model is called \sagrec.
The analysis of the results of these four variants of the model help us to unveil the drivers behind the absence or presence of evolution of the MZR.\\
\indent
Since the MZR involves the oxygen abundance in the cold gas phase, we analyse the mass content of this chemical element in the different baryonic components.
Fig~\ref{fig:OxygenMass} shows the mass of oxygen as a function of stellar mass at redshift $z=0$, $z=2.2$ and $z=3.5$ in the top, middle and bottom panels, respectively, for galaxies generated by the \sagcal~model. 
We separately track the oxygen mass in the hot gas phase (solid line), in the cold gas phase (the thickest solid line), the oxygen mass locked up in stars (the thinnest solid line) and the total oxygen mass ever produced by the stars in  galaxies (dashed line). 
The latter is the sum over time of the oxygen produced by the stars in the galaxy, $\Delta M_{\rm ej}^{\rm O} + \Delta M_{\rm ej(Ia)}^{\rm O}$ (involved in Eqs.~\ref{eq:MhotJ} and \ref{eq:McoldJ}).
Even though the ejected gas component (Eq.~\ref{eq:MejecJ}) does contain some oxygen, we do not show it because it typically is several orders of magnitude lower than the other baryonic components (the mean values of their oxygen mass do not surpass the amount of $\sim 10^{4}$~\msun~at any redshift). 
This responds to the ejected mass component generally being a small contribution to the baryon fraction, except in low mass halos. The latter is a general behaviour of semi-analytic models \citep{Lagos2018}.\\
\indent
Most of the oxygen mass produced by stars is located in the hot gas phase, independently of the stellar mass and redshift. 
However, when time goes by the difference between the oxygen mass in hot and cold gas becomes smaller.
In addition, for galaxies with $M_\star \lesssim 10^{10.0-10.5}$~\msun, the amount of oxygen locked up in stars is lower than in the cold gas phase, which is the  opposite to what we obtain for galaxies of $M_\star \gtrsim 10^{10.0-10.5}$~\msun. 
This inversion takes place in massive galaxies as the mass of their cold gas reservoir becomes much smaller than their stellar mass (see the relation between the cold gas fraction and stellar mass at $z=0$ in Fig.~$3$ of Paper I), which naturally leads to the amount of oxygen contained in cold gas to be smaller than the oxygen locked up in stars.
Because the oxygen in the hot gas dominates the total amount of oxygen in the galaxy-halo system, it is then expected that the exact evolution of the metallicity of the hot gas could have a strong impact on the evolution of a galaxy's metallicity.

\section{The mass-metallicity Relation}
\label{sec:MZR}
We present the current MZR predicted by the model \sagcal~and the evolution of this relation, comparing them with several observational datasets.

\subsection{Mass-metallicity relation at z=0}
\label{sec:MZR0}
\begin{figure}
	\includegraphics[width=1.0\columnwidth]{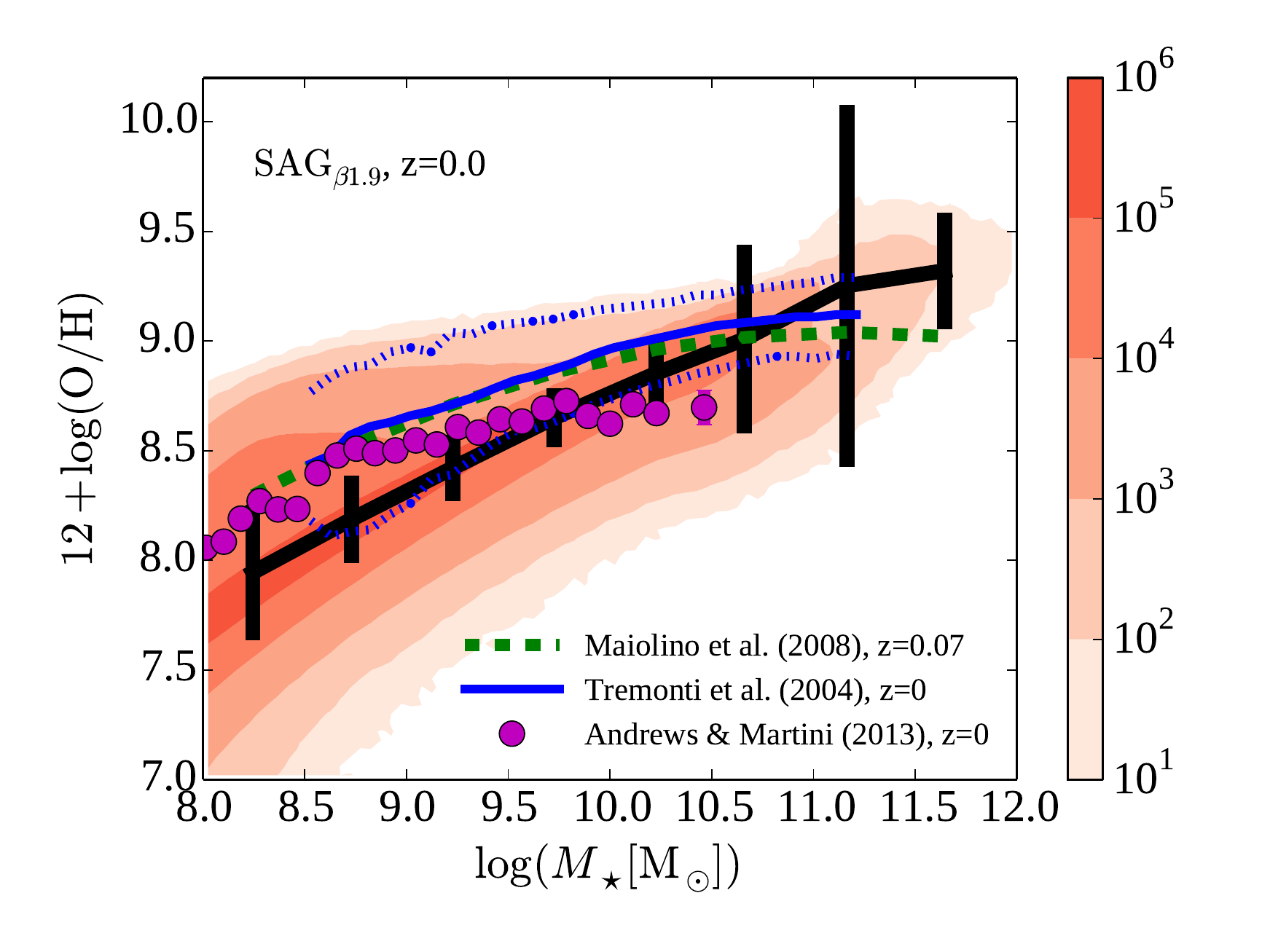}
    \caption{
    MZR of SF galaxies for the model \sagcal~at $z=0$.
The black thick solid line depicts the mean values of the relation with their $1\,\sigma$ dispersion denoted by errorbars. 
The shaded regions show the 2-dimensional histogram of the model galaxies.
We also show the best fit to observations of 
\citet{Maiolino2008} (green dashed line), the mean values and $1\,\sigma$ scatter of observations by \citet{Tremonti2004} (blue solid and dotted lines, respectively), and the mean values from \citet{Martini2013} (filled purple circles).
}
    \label{fig:MZRmodel0}
\end{figure}
Fig.~\ref{fig:MZRmodel0} shows the predicted MZR of star forming (SF) galaxies at $z=0$ in the \sagcal~model (shaded area).
We define SF galaxies as those whose specific SFR (sSFR), i.e., the ratio between the SFR and stellar mass, is higher than $10^{-10.7}\,{\rm yr}^{-1}$ (Paper I), following the criterion adopted in \citet{Brown2017} for the analysis of satellite galaxies selected from the Sloan Digital Sky Survey (SDSS) DR7.

Mean metallicity values for different stellar mass ranges and the corresponding standard deviation are shown with the thick solid line and error bars, respectively.
We compare the prediction of the model with the mean values of the 
spectroscopic observations of \citet{Tremonti2004} and \citet{Martini2013}, and with the fit from \citet{Maiolino2008}.
The latter is calibrated with the spectroscopic observations of $z=0.07$ galaxies by \citet{Savaglio2005}, of $z=2.2$ galaxies by \citet{Erb2006}, and their own spectroscopic observations of $z=3.5$ galaxies. 
This fit is given by
\begin{eqnarray}
        12 + {\rm log(O/H)} = -0.0864 ({\rm log \,M}_{\star} - {\rm log\,M_0})^2 + {\rm K_0},
        \label{eq:Maio}
\end{eqnarray}
\noindent where ${\rm log}\,{\rm M}_0$ and K$_0$ are estimated at each redshift, as listed in their table $5$.
In general, stellar masses in the observations are estimated assuming an IMF that differs from the Chabrier IMF adopted by \sag. 
Both \citet{Tremonti2004} and \cite{Martini2013} adopted a Kroupa IMF \citep{Kroupa2001}. 
To allow a fair comparison with \sag, we shift their stellar masses by $-0.039$~dex \citep{Knebe2017a}.

It is also interesting to stress the different methods and calibrations that were used to measure the cold gas metallicity by these authors. 
\citet{Tremonti2004} use a photoionization theoretical model that takes prominent emission lines to estimate a statistical metallicity. 
\citet{Maiolino2008} also use photoionization models, but only for metallicities $12 + {\rm log(O/H)}>8.35$. 
For lower values of metallicity, they use a method in which the measurements are based on the electron temperature of the gas.
On the other hand, \citet{Martini2013} use a direct method (also based on the electron temperature) for their whole sample. 
As found by \citet{Kewley2008}, the offset on MZR measured with different methods can be as large as $0.7$~dex and may interfere with both the shape and the zero-point of the MZR. 
\citet{Hughes2013} emphasised the importance of using the same method to obtain metallicities in order to avoid the aforementioned discrepancies.

Acknowledging that measuring metallicities is not an easy task, we analyse our model predictions of the MZR at $z=0$. 
\sagcal~produces a MZR with a steeper slope than observed.
At \mstar~$\gtrsim 10^{10}$ \msun, the model MZR agrees with the observations within their errorbars. 
At \mstar~$\lesssim 10^{10}$ \msun, \sagcal~slightly underpredicts the mean metallicities, although still within the uncertainties.

\subsection{Evolution of the Mass-Metallicity Relation}
\label{MZRevol}
\begin{figure}
        \includegraphics[width=1.1\columnwidth]{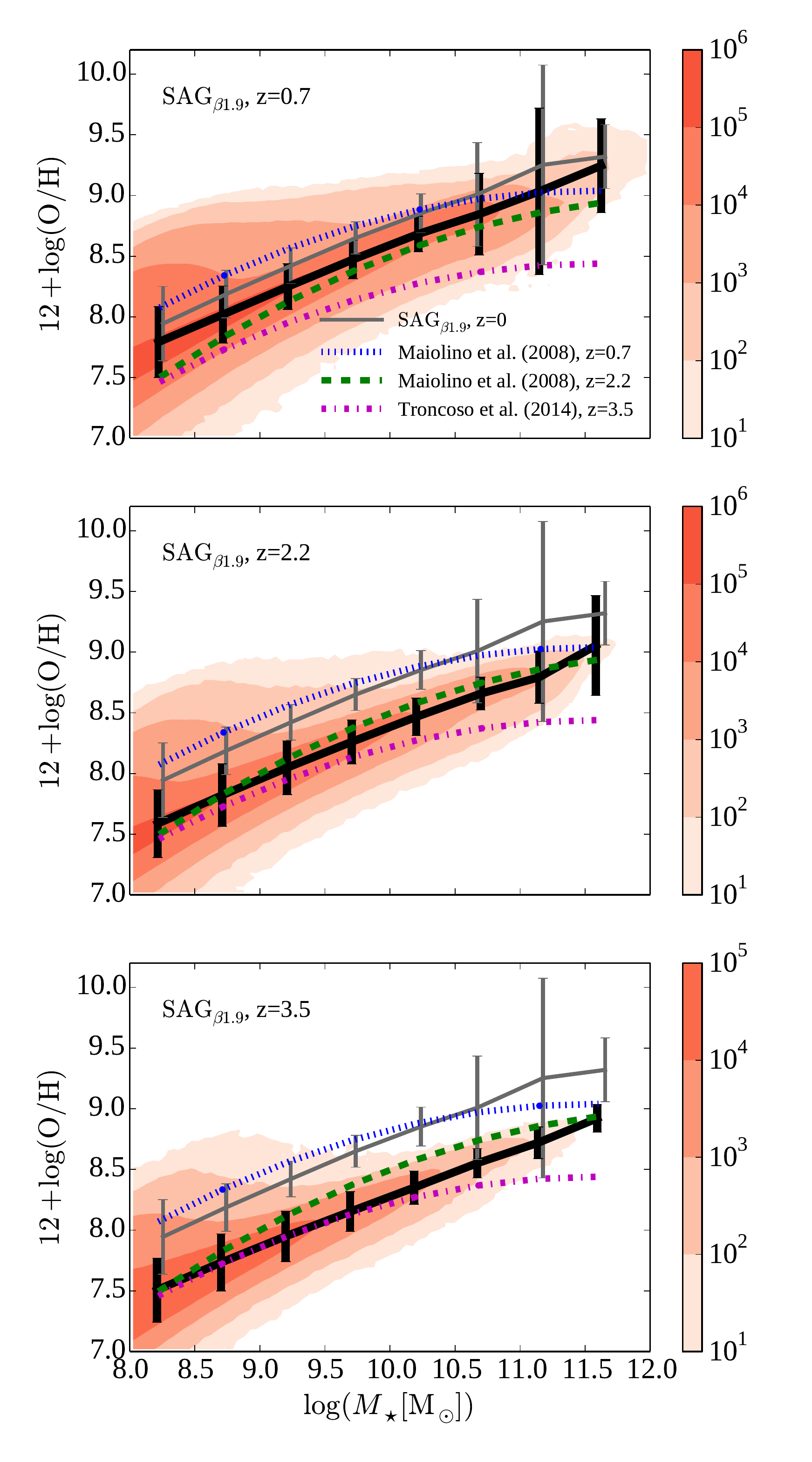}
    \caption{
    MZR of SF galaxies generated by \sagcal~for $z=0.7$ (top panel), $z=2.2$ (middle panel) and $z=3.5$ (bottom panel). 
Shaded regions show the 2-dimensional histogram of the model galaxies, while the black thick solid line represents the mean values of the relation with the corresponding $1\sigma$ scatter shown as errorbars. 
For reference, each panel shows the $z=0$ relation in \sagcal~ as thin solid line. 
In all cases, the best fits to the MZR of \citet{Maiolino2008} at $z=0.7$ and $z=2.2$, and of \citet{Troncoso2014} at $z=3.5$ are shown as blue dotted, green dashed  and purple dashed-dotted lines, respectively.
}
    \label{fig:MZRmodel}
\end{figure}
\begin{figure*}
	\includegraphics[width=1.0\textwidth]{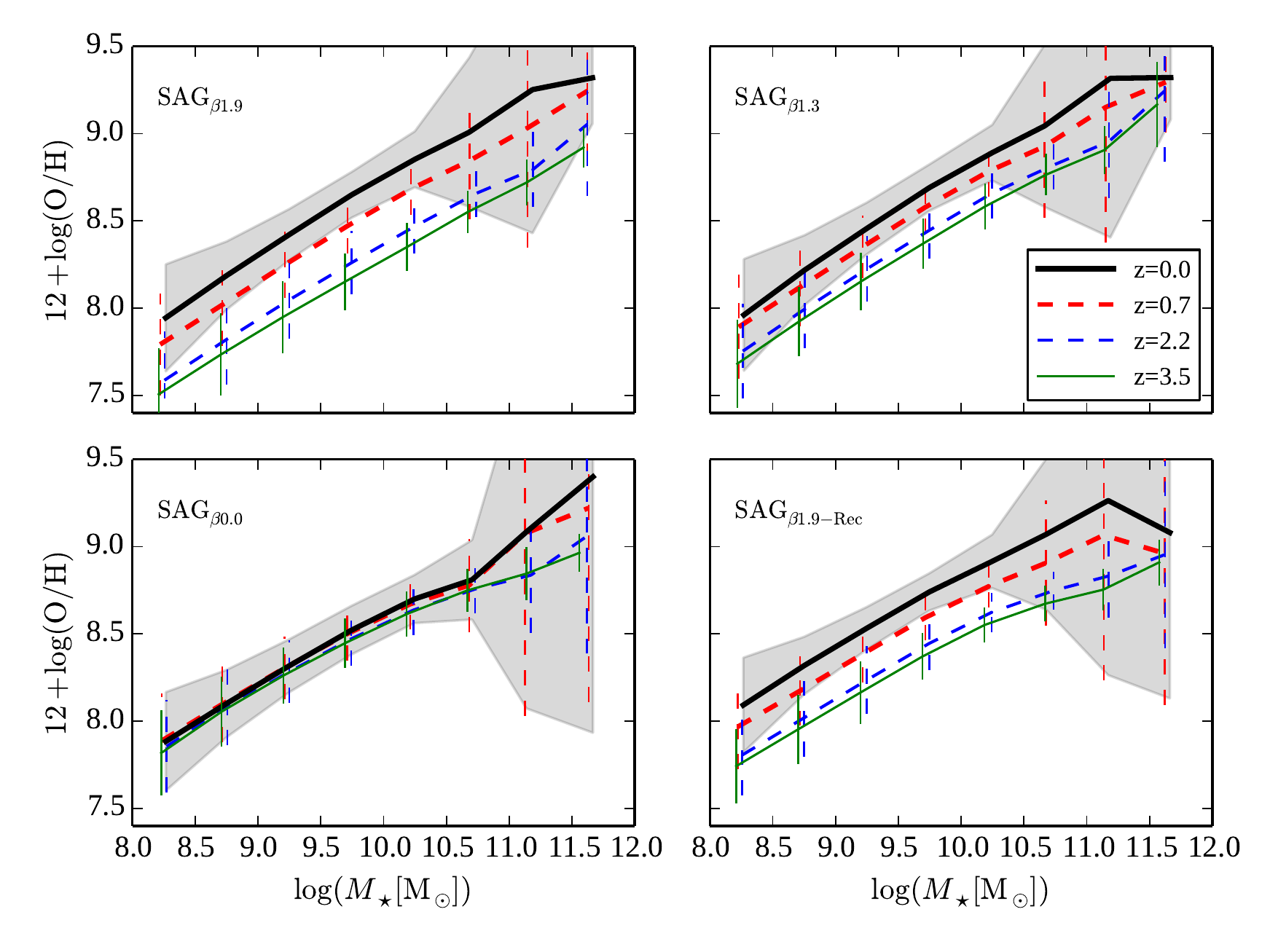}
    \caption{
    MZR of SF galaxies at different redshifts for the \sagcal~(top left panel), \sagbeta~(top right panel), \oldfb~(bottom left panel) and \sagrec~(bottom right panel) models. 
	Different line styles represent mean values of the MZR at
$z=0.0, 0.7, 2.2$ and $3.5$, as indicated in the legend. 
	The corresponding $1\sigma$ scatter is shown as shaded areas for $z=0$ or errorbars, otherwise. 
	The models with the new SN feedback prescription show some level of evolution in their MZR, contrary to the \oldfb~model which shows no evolution at all.
	The evolution of the MZR in the model \sagcal~(top left panel) is of the order of the one found observationally (see Fig.~\ref{fig:MZRmodel}).
	}
    \label{fig:MZRevolution}
\end{figure*}
The MZR of SF galaxies at different epochs is shown in Fig.~\ref{fig:MZRmodel} for model \sagcal~(shaded area). 
The thick solid line denotes the mean metallicity values for each redshift, while the thin solid line depicts the mean values of \sagcal~galaxies at $z=0$, for comparison. 
In both cases, errorbars correspond to the $1 \sigma$ scatter. 
We compare the model MZR with the observational fits from \citet{Maiolino2008} at $z<3$ and from \citet{Troncoso2014} at $z>3$.
The fit introduced by \citet{Maiolino2008} (equation~\ref{eq:Maio}) was later re-calibrated by \citet{Troncoso2014} with the parameters ${\rm log}\,{\rm M}_0$ and K$_0$ obtained from their own spectroscopic observations at $3 < z < 5$ (their metallicity measurements are obtained from the theoretical R$_{23}$ method).
Despite differences in the shape and zero point of the MZR at $z=0$, observations evidence an evolution of this relation. 

At $z=0.7$ (top panel), we can see that there is a tendency of the model MZR to evolve from the relation at $z=0$ towards lower metallicities at fixed stellar mass.
\sagcal~reproduces a relation with a steeper slope (of $\sim 0.40$) than observed, however still within the uncertainties.
On the other hand, the model MZR at $z=2.2$ (middle panel) has a slope of $\sim 0.41$ across the full stellar mass range shown, in good agreement with the observations.
At $z=3.5$ (bottom panel), the model MZR is in good agreement with the observations at \mstar~$\lesssim 10^{10}$ \msun; the slope of both observed and model relations are similar ($\sim 0.41$).
However, for \mstar~$\gtrsim 10^{10}$ the model MZR deviates from the observed one which is characterized by a shallower slope.

Despite the tension between the model and the observations at $z<1$, \sagcal~does reproduce the evolution of the zero-point of the MZR until $z \lesssim 3.5$ (notice that the slopes for all four redshifts are restricted to $\approx 0.40 - 0.41$). 
The redshift evolution of the MZR in SAMs has been reported recently by \citet{Xie2017} but for a more restricted redshift range.
They use an improved of the semi-analytic model \textsc{gaea} \citep{Hirschmann2016}, modified to add a prescription for the molecular gas fraction and the star formation efficiency; the assumed values of model free parameters are also changed. 
These improvements helped find a cosmic evolution of the MZR up to $z \sim 0.7$ for $10^8 \lesssim M_\star \lesssim 10^{12}$~\msun~and up to $z \sim 2$ for $M_\star \gtrsim 10^{10}$~\msun. 
Instead, \sagcal~produces an evolution of the MZR up to $z \sim 3.5$ and $z \sim 2.2$ for $M_\star \lesssim 10^{10.5}$~\msun~and $M_\star \gtrsim 10^{10.5}$~\msun, respectively. 
Additionally, we study how changing the SN feedback modelling and the fate of the recycled material alter the mass and metal content of the different baryonic components affecting the MZR, with the aim of understanding the origin of its evolution.

\section{The origin of the MZR evolution}
\label{sec:Analysis}
The aim of this work is to identify the key physical processes that are responsible for the evolution of the MZR. 
Paper I shows that the SN feedback modelling has a large impact on the evolution of the cosmic SFR and the mass assembly of galaxies, being strongly dependent on the value the parameter $\beta$ takes (Eq.~\ref{eq:feedfire}).
As this mechanism might play an important role in the MZR evolution, we study related properties of the galaxy population generated by the \sagcal, \sagbeta~and \oldfb~models.
The analysis of the \sagrec~model also allows us to evaluate the impact of the fate of the recycled material.

\subsection{Comparison of the MZR predicted by variants of \textsc{SAG}}
\label{sec:CompEvolMZR}
The MZR for SF galaxies at different redshifts predicted by the \sagcal, \sagbeta, \oldfb~and \sagrec~models are presented in the top-left, top-right, bottom-left and bottom-right panels of Fig.~\ref{fig:MZRevolution}, respectively.
The lines show the mean values of the relation for $z=0$ (thick solid line), $z=0.7$ (thick dashed line), $z=2.2$ (thin dashed line) and $z=3.5$ (thin solid line). 
Both the \sagcal~and \sagbeta~models display a clear MZR evolution that is not seen in \oldfb.
The modifications introduced to the SN feedback modelling, described in Sec.~\ref{sec:improve}, drive the evolution of the MZR which is evident from the decrease of its zero-point with increasing  redshift.
Although both \sagcal~and \sagbeta~show evolution in their MZR, it is more pronounced for the former.
Considering the redshift range $0 \leq z \leq 3.5$, the evolution obtained for the normalization of the MZR is $\sim 0.44$~dex for \sagcal~and $\sim 0.28$~dex for \sagbeta.
This difference in the evolution is a consequence of a milder redshift-dependence of the reheated and ejected mass in \sagbeta~with respect to \sagcal. 
On the other hand, the evolution of the MZR is $\sim 0.35$~dex for the \sagrec~model, a bit smaller than for \sagcal, although the features of the SN modelling are the same in both models. 
The change of the fate of the recycled material has a mild impact on the degree of the MZR evolution. 
This result leads us to conclude that the change in the modelling of SN feedback is the main driver of the MZR evolution. 
The way in which the recycled material affects the hot and cold gas phases in model \sagcal~provides a physical channel in which our model is able to reproduce the observed MZR evolution (see Fig.~\ref{fig:MZRmodel}). 
The reason of this is analysed in detail in Section~\ref{sec:quantities}.

The slope of the MZR\footnote{
The slope of the MZR is estimated by applying the linear least squares fitting technique to the stellar mass range $M_{\star} [M_{\odot}]\in [10^8,10^{10.5}]$, where the 1$\sigma$ dispersion is low, as shown in Fig.~\ref{fig:MZRevolution}.
}
does not change significantly over time in any of the models. 
For the redshift range considered, the slopes are $\sim 0.42 - 0.44$, $\sim 0.44 - 0.45$, $\sim 0.39 - 0.38$, and $\sim 0.39 - 0.40$ for \sagcal, \sagbeta, \oldfb~and \sagrec, respectively, where higher values correspond to lower redshifts.
A similar prediction was obtained by \citet{Ma2016} from the analysis of the FIRE simulations, who found that a common slope of $\sim 0.35$ for redshifts in the range $z\in[0,6]$ provides a good fit to their model results. 
Note that the zero-point of their MZR evolves by $\sim 1$~dex within that redshift range, but it is of the order of the one produced by \sagcal~when we restrict to the redshift range analysed in our work.
\citet{Guo2016} compared the predictions of the \textsc{EAGLE} hydrodynamic simulation and two semi-analytic models, \textsc{galform} and \textsc{l-galaxies}, and found that while the former displays a strong evolution in both slope and zero-point of the MZR, the SAMs have a MZR which hardly evolves with redshift. 

\subsection{Evolution of the cosmic density of cold gas and SFR}
\label{sec:coldgas}
The evolution of gas and metal content of the ISM leads to the evolution of the MZR. 
Therefore, we focus on the cosmic density of the cold gas, $\Omega_{\rm cold}$, as a function of redshift, paying special attention to the cosmic density of the hydrogen and oxygen components of this phase ($\Omega_{\rm H}$ and $\Omega_{\rm O}$, respectively).
The gas density is calculated as
\begin{eqnarray}
\Omega_{\rm x} = \frac{\rho_{\rm x}}{\rho_{\rm c (z=0)}} = \frac{M_{\rm x}/L^3}{\rho_{\rm c (z=0)}}
\label{eq:Omega}
\end{eqnarray}
\noindent 
where $\rho_{\rm c (z=0)} = 2.77 \times 10^{11}\,h^2\,M_{\odot}\,{\rm Mpc}^{-3}$ is the critical density at $z=0$ for our adopted cosmology,
and $\rho_{\rm x}$ is the mass density of either the total cold gas (x=cold), or the hydrogen (x=H) and oxygen (x=O) contained in the cold gas. 
The mass density $\rho_{\rm x}$ is estimated by the ratio between the corresponding total mass, $M_{\rm x}$, in the simulation and the cube of the box side-length, $L$, of the underlying cosmological simulation. 

\begin{figure}       
	\includegraphics[width=1.0\columnwidth]{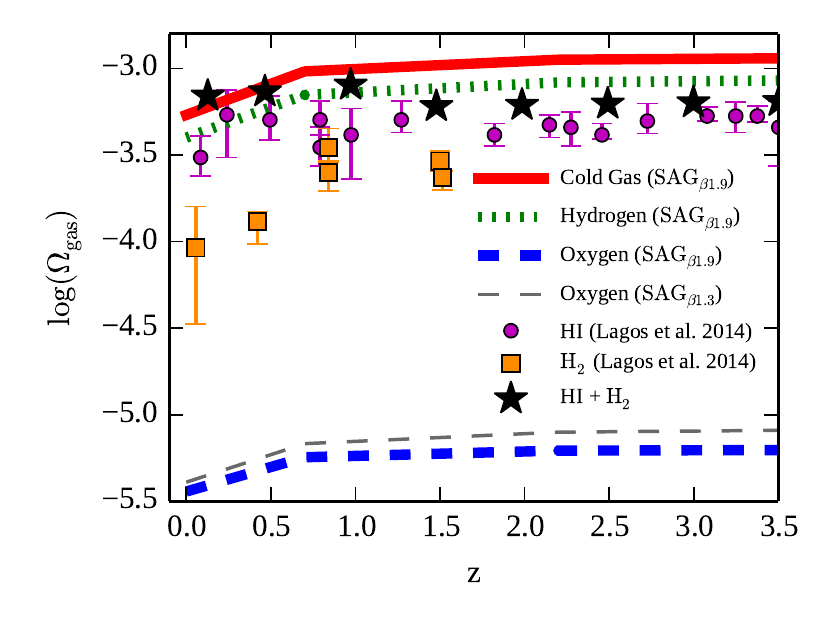}
    \caption{
	Evolution of the cosmic density of the cold gas ($\Omega_{\rm cold}$, red solid line), the hydrogen contained in this phase ($\Omega_{\rm H}$, green dotted line) and the oxygen component ($\Omega_{\rm O}$, thick blue dashed line) for model \sagcal.
	The cosmic density of the oxygen contained in the cod gas 
for model \sagbeta~is represented by a thin grey dashed line. 
	Filled black stars show the observed gas density obtained from the sum of masses of neutral atomic hydrogen HI (filled purple circles) and molecular hydrogen H$_2$ (filled orange squares) presented by \citet{Lagos2014}. 
	The modelled cosmic density of hydrogen shows a good agreement with observations through the redshift range considered ($0 \leq z \leq 3.5$).
	}
    \label{fig:Omegagas}
\end{figure}

Fig.~\ref{fig:Omegagas} shows, for model \sagcal, the cosmic density 
$\Omega_{\rm cold}$ (solid line), $\Omega_{\rm H}$ (dotted line) and $\Omega_{\rm O}$ (thick dashed line) as a function of redshift. 
We compare our results with the observations compiled by \citet{Lagos2014}, in which the gas density of both atomic (HI) and molecular (H$_2$) neutral hydrogen were studied. 
Since the cold gas in the model is not split in these two components, the comparison is made with the observed gas density resulting from the sum of HI and H$_2$. 
For the redshift range considered, $\Omega_{\rm H}$ matches the observed cosmic density of hydrogen, ${\rm HI+H_2}$, reasonable well. 
This agreement is not necessarily surprising as the \sagcal~model has been calibrated to match the cold gas fraction at $z=0$ (see fig.~3 in Paper I), among other constraints (Section~\ref{sec:calibration}). 
The agreement of $\Omega_{\rm H}$ with the observed ${\rm HI+H_2}$ is particularly good at $z \approx 0.5$. 
For lower/higher redshifts, the predicted cosmic density of hydrogen is underestimated/overestimated. 
This is consistent with the discrepancies between model predictions and observations found in the evolution of the cosmic SFR density (SFRD) for model \sagcal, as we show next.
\begin{figure}
 \includegraphics[width=1.0\columnwidth]{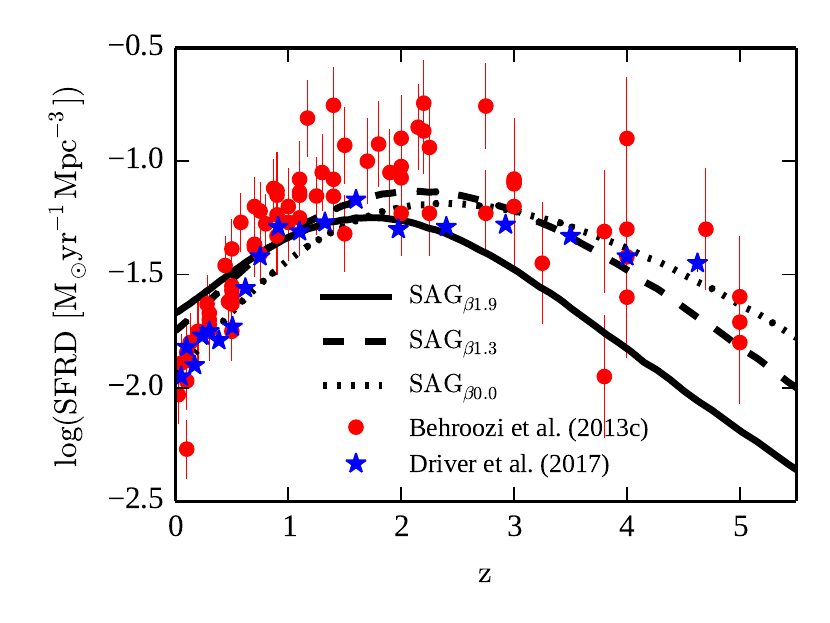}
    \caption{
    Cosmic SFR density as a function of redshift for the models \sagcal~(solid line), \sagbeta~(dashed line) and \oldfb~(dotted line).
Filled circles with error bars show the observations compiled by \citet{Behroozi2013c}, while stars correspond to recent results of \citet{Driver2017}.
In general, all three models agree with the observations. Differences arise from the SN feedback modelling (between \sagcal~and \oldfb) or different values of the parameter $\beta$ associated to the new SN feedback model (between \sagcal~and \sagbeta).
}
    \label{fig:cSFR}
\end{figure}
Fig.~\ref{fig:cSFR} presents the evolution of the cosmic SFRD for the \sagcal, \sagbeta~and~\oldfb~models. 
They are compared with the data compiled by \citet{Behroozi2013c} and the latest results of the combined GAMA/COSMOS/3DHST from \citet{Driver2017}.
The three models show reasonable agreement with the observations, specially at $z<2$, however there are some differences worth discussing. 
The \sagcal~model underpredicts the SFRD at $z \gtrsim 2$, and overpredicts it at $z \lesssim 0.5$.
This differences explain the aforementioned deviations from the observed values of $\Omega_{\rm H}$ (Fig.~\ref{fig:Omegagas}), since star formation takes place from the cold gas reservoir.

The agreement between model predictions and observations improves, particularly at high redshifts, in the \sagbeta~model.
This is expected since the parameter $\beta$, which regulates the redshift dependence of the reheated and ejected mass (Eq.~\ref{eq:feedfire} and~\ref{eq:energySN}, respectively), is fixed to a lower value in \sagbeta~compared to~\sagcal.
Therefore, the weaker effect of SN feedback at higher redshifts in the \sagbeta~model leaves more cold gas available for star formation, leading to higher SFRD in comparison with \sagcal. 
This effect is exacerbated in the \oldfb~model, where gas reheating is equally efficient at fixed SFR and circular velocity for any redshift, in contrast with the evolving efficiency of \sagbeta~and \sagcal, allowing significantly more SF at high redshift.

\begin{figure*}
        \includegraphics[width=1.0\textwidth]{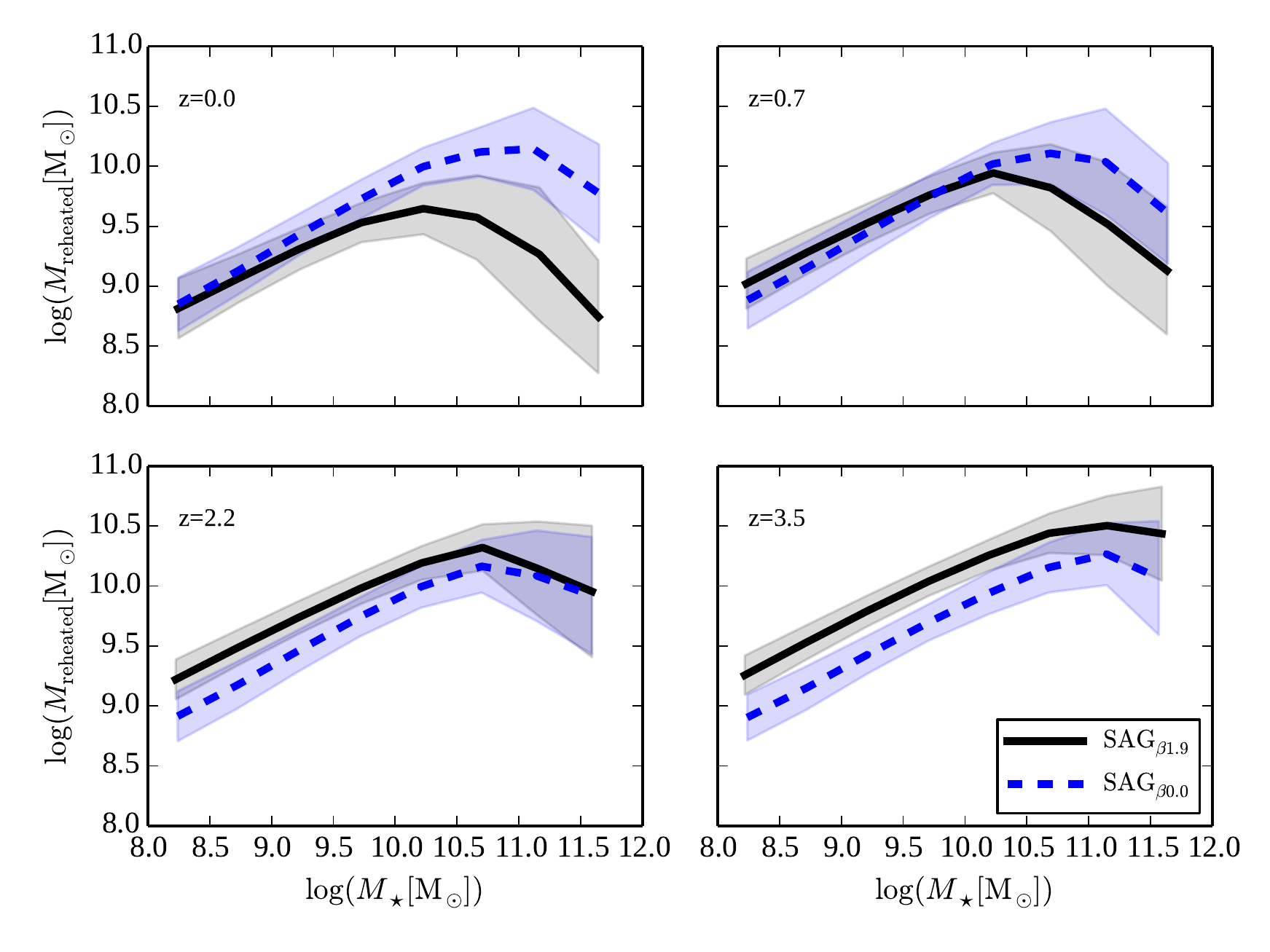}
    \caption{
Reheated mass of SF galaxies as a function of stellar mass at $z=0$ (top-left panel), $z=0.7$ (top-right panel), $z=2.2$ (bottom-left panel) and $z=3.5$ (bottom-right panel).
Black solid and blue dashed lines show the \sagcal~and \oldfb~models, respectively, while the black and blue shaded regions show the corresponding $1\sigma$ scatter.
}
    \label{fig:ReheatMstar}
\end{figure*}

Fig.~\ref{fig:ReheatMstar} shows the mean values of reheated mass as a function of stellar mass for central galaxies at redshifts $z=0$, $z=0.7$, $z=2.2$ and $z=3.5$ (top-left, top-right, bottom-left and bottom-right panels, respectively) for the \sagcal~(solid line) and \oldfb~(dashed line) models; shaded regions show the corresponding $1 \sigma$ dispersion.
We focus on the \sagcal~and \oldfb~models because they produce the largest and the smallest evolution of the zero-point of the MZR, respectively.
We can see that the amount of reheated mass, at fixed stellar mass, shows a mild dependence with redshift in the \oldfb~model.
On the contrary, \sagcal~shows that, at fixed stellar mass, the amount of reheated mass decreases strongly with decreasing redshift. 
This is expected from the explicit redshift dependence introduced in the reheated mass produced by the new feedback model (Eq.~\ref{eq:feedfire}).
This results in higher amounts of reheated mass than \oldfb~at $z\gtrsim 2.2$ and lower amounts at lower redshifts.
\begin{figure*}
        \includegraphics[width=1.0\textwidth]{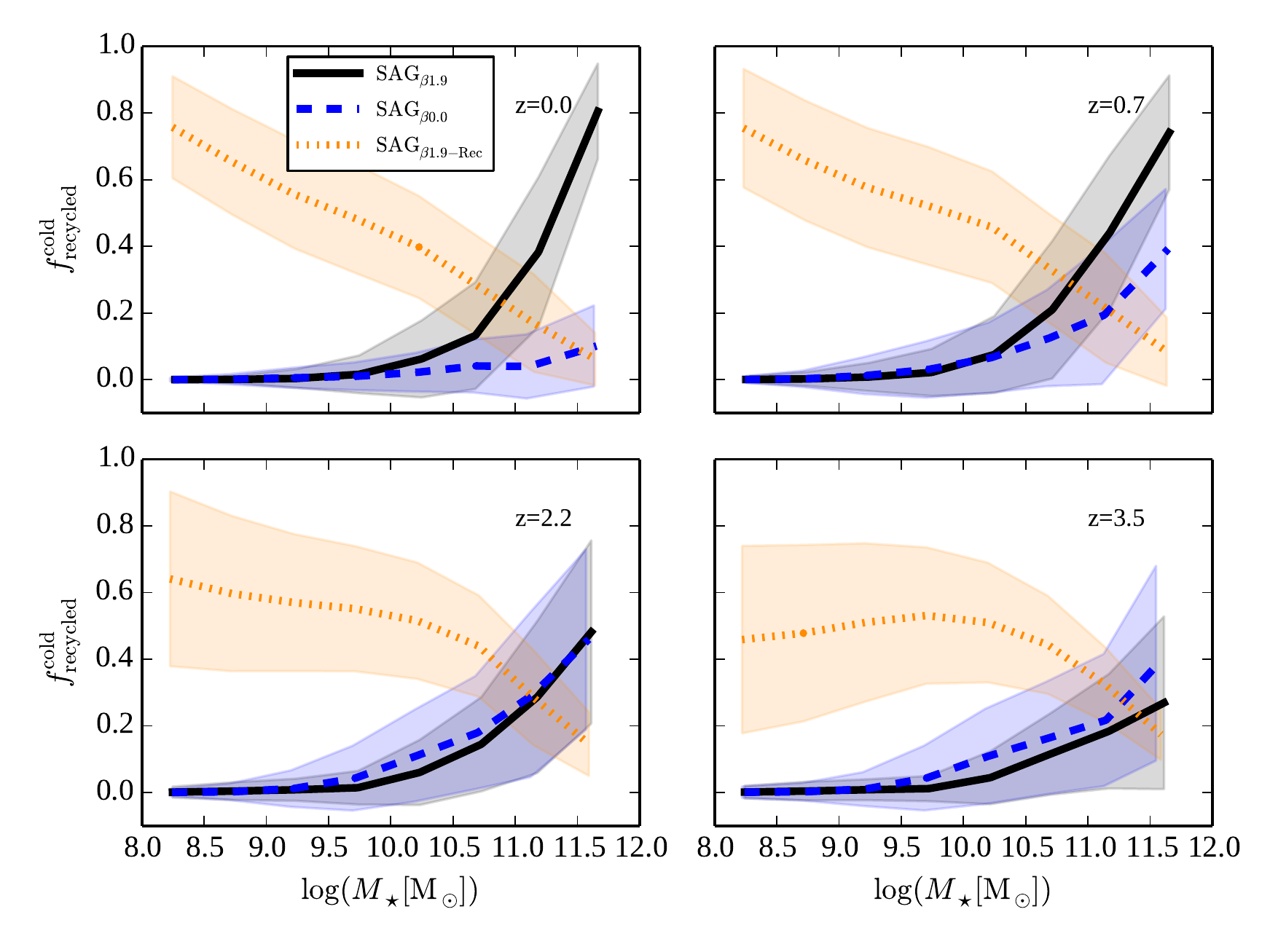}
    \caption{
	Fraction of recycled mass that is transferred directly to the cold gas phase with respect to the total recycled mass ($f_{\rm recycled}^{\rm cold}=M_{\rm recycled}^{\rm cold}/M_{\rm recycled}$), as a function of stellar mass for SF galaxies at $z=0$ (top-left panel), $z=0.7$ (top-right panel), $z=2.2$ (bottom-left panel) and $z=3.5$ (bottom-right panel).
	Black solid, blue dashed and orange dotted lines show results for the \sagcal, \oldfb~and \sagrec~models, respectively, while the black, blue and orange shaded regions represent the corresponding $1\sigma$ scatter.
}
    \label{fig:RecycledMstar}
\end{figure*}
The lower values of SFRD at higher redshifts in the \sagcal~model imply that the amount of metals injected by evolved stars onto the ISM is smaller than in \sagbeta~and \oldfb. 
Indeed, the cosmic density of oxygen is lower in \sagcal~than in~\sagbeta~at all redshifts; the latter is represented by a thin dashed line in Fig.~\ref{fig:Omegagas}.
Besides, the rate of metal enrichment also differs as shown by the different redshift dependence of $\Omega_{\rm O}$ in these models. 
We quantify this by estimating the slopes of the $\Omega_{\rm o}-z$ relation for redshift ranges below and above $z\approx 0.7$, where the relation shows a break. 
For $z \gtrsim 0.7$, these slopes reach values of $\approx 0.015$ and $\approx 0.028$  for the \sagcal~and \sagbeta~models, respectively. 
They become larger for $z \lesssim 0.7$, being $\approx 0.28$ and $\approx 0.32$, respectively. 
Applying the same analysis to $\Omega_{\rm H}$, the slopes are $\approx 0.04$ for $z \gtrsim 0.7$ and $\approx 0.35$ otherwise, for both \sagcal~and \sagbeta; the cosmic densities $\Omega_{\rm cold}$ and $\Omega_{\rm H}$ are pretty similar in these two models and thus, for clarity, we only show the latter for \sagbeta. 
We can see that, for a given redshift range, the difference between the slopes of $\Omega_{\rm O}-z$ and $\Omega_{\rm H}-z$ relations are larger for \sagcal~than for \sagbeta, which indicates that the cold gas is contaminated in a more gradual way when the efficiency of SN feedback is larger at higher redshifts. 
This fact is translated in an appreciable evolution of the MZR.
This analysis supports the justification of the very little evolution of the MZR predicted by the SAM variants discussed by \citet{Hirschmann2016}, which is attributed to the high SFR and consequent rapid chemical enrichment of the cold gas phase at high redshift, which results from the rather low reheating and ejection rates at early times.

Thus, the decrease of the degree of evolution of the zero-point of the MZR is directly linked to the progressive increase in the SFRD at $z\gtrsim 2$ as we reduce the strength of the redshift dependence of the reheated mass from \sagcal~to \oldfb.
SN feedback, therefore affects the evolution of the MZR through the role it has on the regulation of star formation, which in turn determines the amount of reheated gas (equations~\ref{eq:feedbackSN} or~\ref{eq:feedfire}) and the recycled mass containing newly synthesized metals (equations~\ref{eq:MassEjected} and~\ref{eq:EjectedIa}). 
Since the metallicity of the cold gas is given by the relative proportion of mass and metals returned to the ISM, it is the balance between inflows and outflows that transport mass and metals between the different baryonic components (equations~\ref{eq:MhotJ}, \ref{eq:McoldJ} and~\ref{eq:MstarJ}) which helps to determine the degree of evolution of the MZR.

\subsection{Role of metal dilution in outflows}
\label{sec:quantities}
\begin{figure*}
        \includegraphics[width=1.0\textwidth]{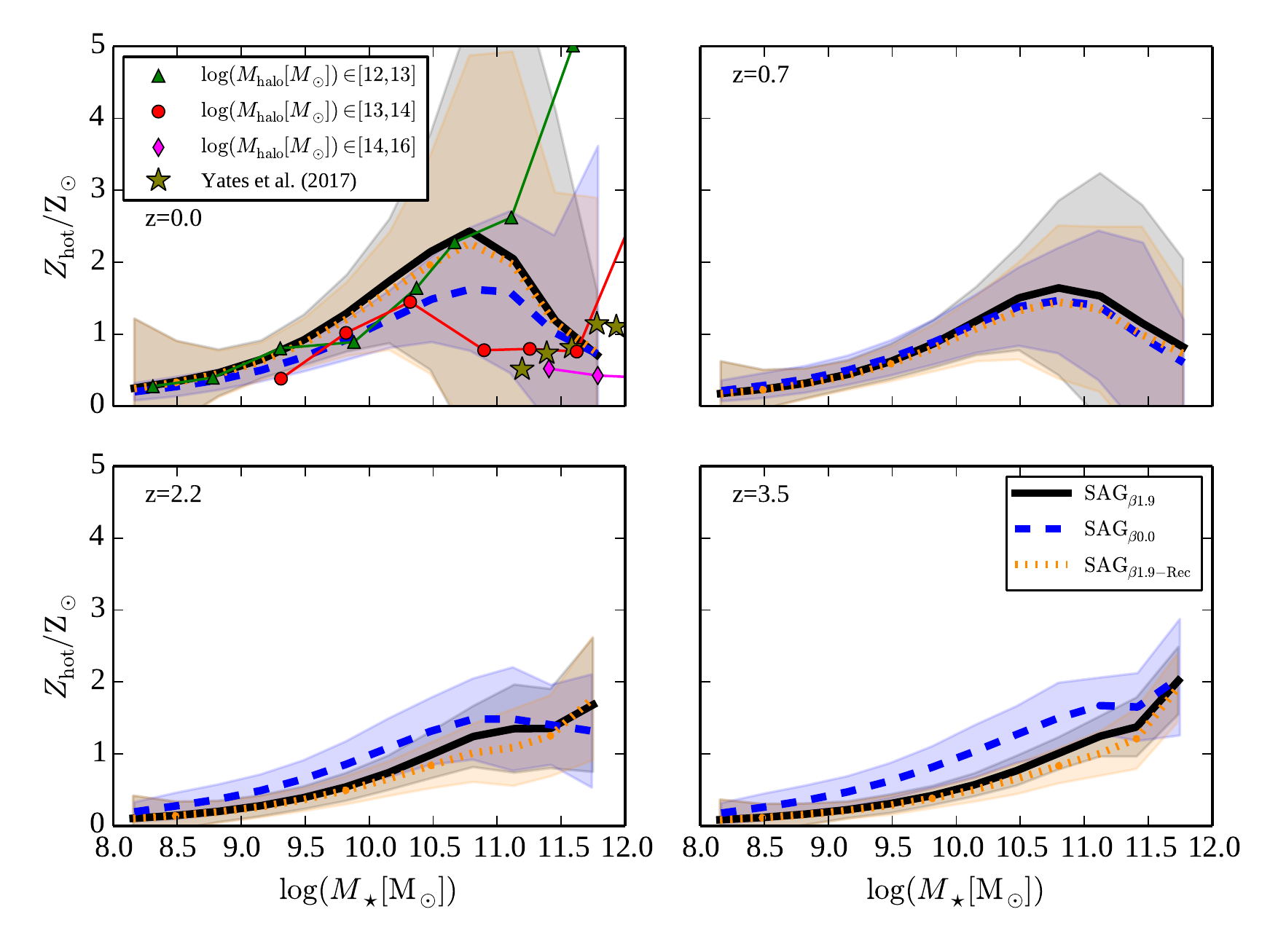}
    \caption{
    Hot gas metallicity $Z_{\rm hot}$ of SF galaxies as a function of stellar mass at $z=0$ (top-left panel), $z=0.7$ (top-right panel), $z=2.2$ (bottom-left panel) and $z=3.5$ (bottom-right panel).
	Black solid, blue dashed and orange dotted lines show the mean values for the \sagcal, \oldfb~and \sagrec~models, 
respectively, while the black, blue and orange shaded regions show the corresponding $1\sigma$ scatter. 
	In the top-left panel, the median hot gas metallicity is estimated for galaxies generated by \sagcal~binned in ${\rm M_{halo}}$ ranges: ${\rm M_{halo} [{\rm M_\odot}]} \in [10^{12}, 10^{13}]$ (green triangles), ${\rm M_{halo} [{\rm M_\odot}]} \in [10^{13}, 10^{14}] $ (red circles) and ${\rm M_{halo} [{\rm M_\odot}]} \in [10^{14}, 10^{16}]$ (pink diamonds), which help to interpret the wide dispersion in metallicity at high stellar mass. Stars represent observational results from \citet{Yates2017} after conversion to the $Z_{\rm hot}$-$M_{\star}$ plane.
	}
    \label{fig:ZHotMstar}
\end{figure*}
We have shown so far that the cold gas is contaminated more gradually in time when the SN feedback effect is stronger at higher redshifts, which gives place to a larger evolution of the zero-point of the MZR.
SN explosions not only inject energy into the ISM and intergalactic medium (IGM), but also newly synthesized metals that are contained in the recycled mass.
We have also seen that the fate of the recycled mass plays a part in determining the magnitude of the MZR evolution (see Section~\ref{sec:CompEvolMZR}).
According to the two possible schemes adopted to regulate the mass transfer between baryonic components (see Sec.~\ref{sec:regulation}), the fate of the recycled mass may depend either on its relative proportion with respect to the reheated mass (as in the \sagcal, \sagbeta~and \oldfb~models) or on the cold gas fraction (as in  the \sagrec~model). 
We refer to the former as recycling scheme 1 and to the latter as recycling scheme 2.
We now analyse the way in which these schemes influence the cold gas metallicity.

Fig.~\ref{fig:RecycledMstar} shows the fraction of recycled mass that is transferred directly to the cold gas, $M_{\rm recycled}^{\rm cold}$, with respect to the total recycled mass, $M_{\rm recycled}$, i.e. $f_{\rm recycled}^{\rm cold}=M_{\rm recycled}^{\rm cold}/M_{\rm recycled}$, as a function of stellar mass for SF galaxies at $z=0$ (top-left panel), $z=0.7$ (top-right panel), $z=2.2$ (bottom-left panel) and $z=3.5$ (bottom-right panel), for the \sagcal~(solid line), \oldfb~(dashed line) and \sagrec~(dotted line) models, respectively.
The different schemes adopted to define the fate of the recycled mass produce completely opposite trends. 
The fraction of recycled mass that is deposited in the cold gas phase increases with increasing stellar mass in the recycling scheme 1, while it decreases for more massive galaxies in the recycling scheme 2.\\
\indent
Both the \sagcal~and \oldfb~models, which adopt the recycling scheme 1, have $f_{\rm recycled}^{\rm cold}\approx 0$ for low-mass galaxies ($M_{\star} \lesssim 10^{10}\,M_{\odot}$) because the reheated mass is larger than the recycled mass for those galaxies in any model.
This implies that the newly synthesized metals are injected directly into the hot gas halo. 
For galaxies with $M_{\star} \gtrsim 10^{10}\,M_{\odot}$ at a given redshift, the fraction $f_{\rm recycled}^{\rm cold}$ increases with increasing stellar mass for both models, consistent with the fact that the reheated mass decreases, as can be seen from Fig.~\ref{fig:ReheatMstar}, allowing more recycled mass to be  transferred to the cold gas directly.
For a given stellar mass within the more massive range, the dependence on redshift of the fraction $f_{\rm recycled}^{\rm cold}$ differs between models characterized by different SN feedback. 
While $f_{\rm recycled}^{\rm cold}$ progressively increases when redshift decreases for \sagcal, the opposite behaviour is seen in \oldfb. 
The redshift dependence of the efficiency of SN feedback in the former produces less amount of reheated mass at lower redshift (see Fig.~\ref{fig:ReheatMstar}), so that $\approx 80$ per cent of the recycled mass ends in the cold gas for the most massive galaxies at $z=0$ in \sagcal.\\
\indent
In the recycling scheme 2, $f_{\rm recycled}^{\rm cold} \gtrsim 0.5$ for low-mass galaxies at any redshift. 
This fraction decreases for more massive galaxies, approaching values close to zero for the most massive ones, specially at low redshifts ($z\lesssim 0.7$). 
These trends are a result of the dependence of cold gas fraction on stellar mass, which is higher for less massive galaxies (see fig.~3 in Paper I for the model \sagcal).
Therefore, while low-mass galaxies have cold gas mainly polluted by recycled material that is kept in this reservoir, the low cold gas fraction of massive galaxies produces a larger fraction of recycled material to be ejected into the hot gas halo. 
Thus, the cold gas of high-mass galaxies is chemically enriched mainly through cooled gas. 
Although the gas cooling rate becomes lower for more massive galaxies as a result of AGN feedback, the contaminated cooling flow has an important impact on the chemical enrichment of the cold phase because of the low cold gas fraction of these galaxies.\\
\indent
The pollution of the cold gas takes place through two channels, namely, the direct contamination of the cold gas by recycled material that is directly injected onto this phase (cases with high values of $f_{\rm recycled}^{\rm cold}$), and gas cooling, which contributes with a mass of cooled gas with a metallicity determined by the one achieved by the hot gas halo (cases with low values of $f_{\rm recycled}^{\rm cold}$).
For a given stellar mass, the relative importance of each channel depends on the recycling scheme adopted. 
For the cases dominated by gas cooling, the metallicity of the hot gas is a direct driver of the metallicity of the cold gas. 
Fig.~\ref{fig:ZHotMstar} shows the mean value of the hot gas metallicity, $Z_{\rm hot}$, as a function of stellar mass for SF galaxies in the \sagcal~(solid line), \oldfb~(dashed line) and \sagrec~models, at different redshifts.
The metallicity $Z_{\rm hot}$ is defined as the ratio between the mass of metals contained in the hot gas phase and its total mass; it is normalized by the solar value of $Z_{\odot} = 0.0196$ \citep{VonSteiger2016}. 
In all models, galaxies with higher stellar mass are characterized by more chemically enriched hot gas, except for galaxies with $M_{\star}\gtrsim 5 \times 10^{10}\,{\rm M}_{\odot}$ at low redshifts ($z \lesssim 0.7$), for which the trend is inverted. 
Besides, the dispersion of $Z_{\rm hot}$ becomes quite large for this stellar mass range, particularly for model \sagcal.\\
\indent
Focusing on the way the hot gas increases its metallicity, it can be seen that, regardless of stellar mass, the hot gas of SF galaxies in the \oldfb~model is more chemically enriched than the hot component of those galaxies in \sagcal~at high redshifts ($z \gtrsim 2.2$). 
As the redshift decreases, there is a reversal of the trend, with \sagcal~displaying a more polluted hot gas than \oldfb. 
Such a reversal is a result of the different rate of chemical enrichment of the hot gas halo in these two models as the stellar mass of the associated galaxy grows. 
In \sagcal, the metallicity of the hot gas increases more gradually than in \oldfb~because of the progressive decrease of the reheated mass (outflows) with decreasing redshift in the former model,
specially for high-mass galaxies (see Fig.~\ref{fig:ReheatMstar}), which gives rise to higher values of $f_{\rm recycled}^{\rm cold}$ (see Fig.~\ref{fig:RecycledMstar}).
Therefore, in \sagcal, the recycled and newly synthesized metals that are injected into the hot gas halo through outflows are less diluted in the mass carried by them as redshift decreases.
Moreover, the more gradual increase of $Z_{\rm hot}$ in \sagcal~is favoured by lower levels of SF and, consequently, lower mass of metals produced at higher redshifts with respect to either \sagbeta~or \oldfb~(Figs.~\ref{fig:Omegagas} and \ref{fig:cSFR}).
The hot gas in model \sagrec~reaches levels of enrichment only slightly lower than in \sagcal, with pretty similar dependence on stellar mass and redshift. This small difference is  
consistent with the minor direct contribution of recycled material received by this gas phase. 
In any case, most of the oxygen produced by stars is deposited in the hot gas, as shown by Fig.~\ref{fig:OxygenMass} for \sagcal.\\
\indent
All this analysis helps to disentangle the key physical aspects in the evolution of the MZR.
On one hand, we have two models that differ in the modelling of SN feedback but have the same recycling scheme, i.e. \sagcal~and \oldfb. 
The former produces a zero-point evolution consistent with observations, while the latter does not show any evolution at all. 
Both of them present a fraction $f_{\rm recycled}^{\rm cold}$ that departs from zero only for high-mass galaxies, pointing to the fact that the contamination of the cold gas is mainly driven by gas cooling. 
Hence, the key aspect that explains the evolution of the MZR is the rate at which the hot gas halo is chemically enriched. 
Both the mass of outflows and metals produced are a direct consequence of the action of SN feedback.

On the other hand, the strong change in the trend of the $f_{\rm recycled}^{\rm cold}-M_{\star}$ relation when the recycling scheme is modified has little effect on the evolution of the zero-point of the MZR. 
Since this evolution is larger when the recycling scheme 1 is adopted ($\approx 0.44$ for \sagcal) than when the recycling scheme 2 is implemented ($\approx 0.35$ for \sagrec), we can assert that the different degree of evolution between \sagcal~and \sagrec~is produced by the faster rate of chemical enrichment suffered by the cold gas in the latter as a result of the high values of $f_{\rm recycled}^{\rm cold}$, specially for low-mass galaxies, which reach higher values of cold gas metallicity. 
For high-mass galaxies, gas cooling gains relevance and the lower values of $Z_{\rm hot}$ in \sagrec~help to recover the flattening of the high-mass end of the MZR (see bottom-right panel of Fig.~\ref{fig:MZRevolution}), in better agreement with observations.

\subsubsection{On the features of the $Z_{\rm hot}$-$M_{\star}$ relation}
\label{sec:ZhotDispersion}
The relation between the hot gas metallicity and the stellar mass is characterized by an inverted trend at high stellar masses for $z \lesssim 2.2$ and a quite large scatter compared to what is expected for the ICM.
With the aim of understanding these features of the $Z_{\rm hot}$-$M_{\star}$ relation, we estimate the median values of $Z_{\rm hot}$ of SF galaxies in \sagcal~binned according to the mass of the DM (sub)halo, $M_{\rm halo}$, they reside in.
$M_{\rm halo}$ is given by the mass of each (sub)structure contained within $r_{200}$, which is the radius where the matter density is $200$ times that of the critical density of the universe, i.e., ${\rm M_{halo}} = {\rm M_{DM}(r_{200})}$. 
The top-left panel of Fig.~\ref{fig:ZHotMstar} shows the results for three ranges of halo mass, namely, $M_{\rm halo}{\rm [M_\odot]} \in [10^{12}, 10^{13}]$, $[10^{13}, 10^{14}]$ and $[10^{14}, 10^{16}]$.
As it can be seen, the hot gas metallicity of galaxies inhabiting haloes with lower mass is higher for high-mass galaxies, producing a large dispersion.
While galaxies in the lowest halo mass bin drive the increase of the mean value of $Z_{\rm hot}$ with stellar mass, those galaxies with $M_{\star}\gtrsim 5 \times 10^{10}\,{\rm M}_{\odot}$ residing in the two most massive halo mass bins are responsible for the decay of $Z_{\rm hot}$. 
Thus, both the scatter and the inversion of this relation are strongly correlated with the halo mass.

The hot gas of central galaxies residing in the two more massive halo mass bins represent the intragroup/intracluster medium (ICM). 
It is found that the galaxy population with $M_{\star}\gtrsim 10^{11}\,{\rm M}_{\odot}$ is dominated by centrals.
Indeed, for the haloes within the mass ranges $M_{\rm halo}{\rm [M_\odot]} \in [10^{12}, 10^{13}]$, $[10^{13}, 10^{14}]$ and $[10^{14}, 10^{16}]$, 
the corresponding percentage of central galaxies
with respect to the total number of galaxies (centrals plus satellites) selected according to this stellar mass cut is $\approx 88$, $94$ and $97$ per cent. 
Thus, the mean values of $Z_{\rm hot}$ for different halo mass bins can be compared with the observed metallicity of hot gas surrounding galaxy groups and clusters presented by \citet[][stars in the top-left panel of Fig.~\ref{fig:ZHotMstar}]{Yates2017}, even though the whole sample of model galaxies includes also satellite galaxies.
These authors show the relationship between the mass-weighted iron abundance within $r_{500}$, ${\bar Z}_{\rm Fe, 500}$ and the hot gas temperature ${T_{500}}=T(r_{500})$ of the ICM for a homogenized dataset of groups and clusters compiled from the literature, where $r_{500}$ is analogous to $r_{200}$ but the matter density is 500 times that of the critical density.
The conversion of ${\bar Z}_{\rm Fe,500}$-$T_{\rm 500}$ to $Z_{\rm hot} - M_{\star}$ is explained in the Appendix~\ref{ap:conversion}.
Considering galaxies within haloes of mass $M_{\rm halo} > 10^{13} {\rm M_\odot}$, which is the mass range considered by \citet{Yates2017}, we can see that the hot gas metallicity predicted by the \sagcal~model is in good agreement with observational data. 
However, model galaxies with $M_{\star} \gtrsim 10^{11}\,{\rm M}_{\odot}$ within haloes of mass $M_{\rm halo}{\rm [M_\odot]} \in [10^{14}, 10^{16}]$ show an opposite trend, with decreasing $Z_{\rm hot}$ for increasing stellar mass.  
We understand this behaviour as due to the SAG model producing an artificially high merger rate in these galaxies at $z<2$ (discussed in detail in Paper I). 
Because of this, the hot haloes of central galaxies acquire too much low metallicity gas from the lower mass galaxies that are merging, resulting in metallicity dilution. 
Similarly, \citet{Yates2017} point out that a major process diluting the hot gas metallicity is the infall of satellites in massive systems, which favours accretion of pristine or low metallicity gas.

\section{Conclusions}
\label{sec:Conclusions}
We have analysed the MZR predictions of the latest version of the semi-analytic model of galaxy formation \sag, described in detail in Paper I. 
The latest improvements implemented in \sag~include a robust model of environmental effects through the action of RPS and TS coupled to the integration of the orbits of orphan satellites, and a higher efficiency of SN feedback at early times allowed by an explicit redshift dependence of the reheated and ejected mass. 
The latter is inspired on the results of the hydrodynamical simulation suite FIRE \citep{Muratov2015}. 
Paper I shows that a variant of the model referred to as \sagcal~allows us to achieve good agreement with observational results for a large range of galaxy properties locally and at high redshift; the SMF at $z=0$ and $z=2$, the SFR function, the scaling relations between black hole and bulge mass, between the cold gas fraction and stellar mass, between the stellar and host halo mass, between the atomic gas content and stellar mass and the evolution of the SFRD. 
In particular, the detailed non-instantaneous chemical enrichment modelling makes \sag~ideally suited to carry out the present work.

Reproducing the development of the MZR has been a long standing problem for SAMs.
The importance of the study of the origin of this relation and the way in which it evolves resides in the insights that it provides on the chemical enrichment of the cold and hot gas phases in connection with gas inflows and outflows.

We summarize our conclusions as follows:
\begin{itemize}
\item 
A SN feedback model which gives rise to larger amounts of reheated and ejected mass at higher redshifts allows us to reproduce the observed evolution of the MZR normalization.
This evolution is found for a wide range of stellar masses ($10^{8}$ \msun~$\lesssim$ \mstar~$\lesssim 10^{12}$ \msun) until $z \lesssim 3.5$ for the model \sagcal, in which the new SN feedback recipe involves an explicit redshift dependence (Fig.~\ref{fig:MZRmodel0} and \ref{fig:MZRmodel}). 
The lack of such dependence, as in the old SN feedback scheme, i.e. the model \oldfb, prevents the MZR normalization from evolving (Fig.~\ref{fig:MZRevolution}).
We also find that the new SN feedback scheme yields no evolution of the slope of the MZR.
\item 
We experimented with the redshift dependence of the reheated and ejected mass in the new SN feedback recipe (represented by the parameter $\beta$) and found that a larger $\beta$ directly translates into a stronger evolution of the MZR normalization ($\approx 0.44$~dex). 
While the \sagbeta~model, which has $\beta = 1.3$, produces some evolution of the MZR normalization ($\approx 0.28$~dex), it is not as pronounced as the one achieved by the \sagcal~model, which adopted $\beta = 1.99$ (Fig.~\ref{fig:MZRevolution}).
Higher values of the parameter $\beta$ embody more violent effects of SN feedback at high redshifts.
\item 
Larger feedback effects at higher redshifts produce larger discrepancies between the modelled and observed SFRD, with the former being below recent data by a factor $\approx 3$ (Fig.~\ref{fig:cSFR}).
Models with stronger SN feedback effect at higher redshifts produce lower values of SFR, with the consequence of a lower production of metals, as indicated by the cosmic density of oxygen content of the cold gas (Fig.~\ref{fig:Omegagas}).
\item
Stronger SN feedback at higher redshifts leads to a delay in the pollution of the hot gas.
For a given stellar mass, ${\rm Z_{hot}}$ increases with decreasing redshift in the model \sagcal, in contrast with the model \oldfb, in which ${\rm Z_{hot}}$ practically remains the same throughout time at fixed stellar mass (Fig.~\ref{fig:ZHotMstar}). 
The lower levels of metal enrichment of the hot has halo at higher redshifts in \sagcal~are a result of the lower values of SFR and the larger amounts of reheated mass which dilute the metals injected in that gas phase (Fig.~\ref{fig:ReheatMstar}).
The slower increase of the hot gas metallicity compared with the stellar mass build-up is finally translated as an evolution of the MZR normalization, since part of the cold gas contamination takes place through 
gas cooling. 
\item
In our model, the metal pollution of the cold gas occurs either through direct injection of recycled material into the cold gas or by the gas cooling of enriched hot gas.
For a given stellar mass, the relevance of each process depends on the criterion (recycling scheme) adopted to decide the fate of the recycled material, i.e. cold or hot gas phase.
When the fraction of recycled mass that ends up in the cold gas phase is defined by the ratio between the reheated and recycled mass (as in the \sagcal, \sagbeta~and \oldfb~models), 
gas cooling 
is the main channel of pollution for low-mass galaxies ($M_{\star} \lesssim 10^{10}\,M_{\odot}$) but losses importance as stellar mass increases.
The opposite trend is obtained when the fate of the recycled mass is regulated by the cold gas fraction of galaxies (Fig.~\ref{fig:RecycledMstar}).
\item 
The modelling of the metal loading of the outflows does not affect our conclusions.
An experiment in which the fate of the recycled material is regulated by the cold gas fraction (model \sagrec, recycling scheme 2), instead of being driven by the amount of reheated mass as in \sagcal~(recycling scheme 1), shows that the galaxy population exhibits an evolution of the MZR normalization that is only slightly smaller ($\approx 0.35$~dex) than  our original model (Fig.~\ref{fig:MZRevolution}). 
This small difference is attributed to a faster enrichment of the cold gas in \sagrec~because of the predominant role of direct metal injection onto this gas phase for low-mass galaxies.
\end{itemize}

Overall, with SAG we find that a stronger evolution of the SN feedback effect is key in getting the expected evolution of the MZR normalization, which has been a long standing challenge for SAMs. 
Both the mass of outflows and metals produced are a direct consequence of the action of SN feedback.
Thus, the evolution of the zero-point of the MZR is mainly caused by the lower levels of star formation, and consequent lower metal production, as a result of stronger SN feedback at high redshift, with minor dependence on the fate of the metals produced by stellar evolution and returned via stellar winds and SNe. 
The explicit redshift dependence involved in the estimate of outflows generated by SN feedback is interpreted as evolving aspects of the ISM not captured by the model, such as different redshift evolution of local disc properties with respect to the circular velocity of galaxies \citep{Lagos2013, Creasey2013}, or additional sources of energy (other than SNe) that may play an important role \citep{Hopkins2012}.
%

\section*{Acknowledgements}
We thank the referee for the constructive comments provided that contribute to improve this manuscript. 
We also thank Lisa Kewley for fruitful discussions and suggestions. 
The authors gratefully acknowledge the Gauss Centre for Supercomputing e.V. (www.gauss-centre.eu) and the Partnership for Advanced Supercomputing in Europe (PRACE, www.prace-ri.eu) for funding the \textsc{MultiDark} simulation project by providing computing time on the GCS Supercomputer SuperMUC at Leibniz Supercomputing Centre (LRZ, www.lrz.de). 
The MDPL2 simulation has been performed under grant pr87yi. 
This work was done in part using the Geryon computer at the Center for Astro-Engineering UC, part of the BASAL PFB-06, which received additional funding from QUIMAL 130008 and Fondequip AIC-57 for upgrades.
FC, and CVM acknowledge CONICET, Argentina, for their supporting fellowships.
SAC acknowledges funding from {\it Consejo Nacional de Investigaciones Cient\'{\i}ficas y T\'ecnicas} (CONICET, PIP-0387), {\it Agencia Nacional de Promoci\'on Cient\'ifica y Tecnol\'ogica} (ANPCyT, PICT-2013-0317), and {\it Universidad Nacional de La Plata} (G11-124), Argentina.
CL is funded by an Australian Research Council Discovery Early Career Researcher Award (DE150100618) and by the Australian Research Council Centre of Excellence for All Sky Astrophysics in 3 Dimensions (ASTRO 3D), through project number CE170100013. The Cosmic Dawn Centre is funded by the Danish National Research Foundation.

\bibliographystyle{mnras}
\bibliography{collacchioni}


\appendix
\section{Conversion of ${\bar Z}_{\rm Fe,500}$-$T_{\rm 500}$ to 
$Z_{\rm hot}$-\mstar}
\label{ap:conversion}
Mean values of the hot gas metallicity $Z_{\rm hot}$ of galaxies in \sagcal\ are presented as a function of stellar
mass in Fig.~\ref{fig:ZHotMstar} (top-left panel). 
Model results are compared with the observed iron abundances ${\bar Z}_{\rm Fe,500}$\footnote{
$ \log(Z_{\rm Fe}) \equiv \left[\frac{\rm Fe}{\rm H}\right] = \log{\left(\frac{N_{\rm Fe}}{N_{\rm H}}\right)} - \log{\left(\frac{N_{\rm Fe,\odot}}{N_{\rm H,\odot}}\right)}$, with the solar iron abundance being $\left(\frac{N_{\rm Fe,\odot}}{N_{\rm H,\odot}}\right)=3.16\times 10^{-5}$ \citep{Grevesse1998}. 
}
of hot gas surrounding groups and clusters presented by \citet{Yates2017}.
Since their iron abundances are given as a function of the hot gas temperature ${T_{\rm 500}}$, we have to make two conversions based on model results.  

On one hand, to convert from ${\bar Z}_{\rm Fe,500}$ to $Z_{\rm hot}$ in the observations, we apply the correlation between the hot gas metallicity and the iron abundance of the hot gas that we obtain internally in \sagcal. 
This is ${Z_{\rm hot}} = 1.55 \, \log(Z_{\rm Fe, SAG}) + 1.92$, which already takes into account the different solar values used to normalize each of these quantities. 
We recall that, in the model, $Z_{\rm hot}$ has been normalized to the solar value of Z$_\odot = 0.0196$ \citep{VonSteiger2016}. 
The iron abundance $Z_{\rm Fe,SAG}$ is referred to the same solar value adopted by \citet{Yates2017}.
We then use observed values of ${\bar Z}_{\rm Fe,500}$ instead of $Z_{\rm Fe, SAG}$ as an entry in this relation to obtain the corresponding $Z_{\rm hot}$. 

On the other hand, we obtain a relation between stellar mass and and gas temperature at $r_{500}$ from model galaxies.
This radius is obtained from $r_{200}$ (given in the simulation outputs) by adopting an isothermal density profile.
The hot gas temperature of model galaxies, ${T_{\rm 500, SAG}}$, is estimated following the procedure described in \citet[][see their sec. 2.3]{Yates2017}, in which the temperature at $r_{200}$, ${T_{\rm 200}}$, is needed. 
This property is calculated by applying the conversion between virial velocity and temperature from \citet{Springel2001}, 
\begin{eqnarray}
T_{200} = 35.9 \, (V_{\rm vir}/{\rm km\,s^{-1}})^2 \, K,
\end{eqnarray}
\noindent and plays a key role in the classification of galaxy environments \citep{Yates2017}. 
Groups are those haloes with temperature $kT_{200} < 2.5\,{\rm kev}$ ($k$ being the Boltzmann's constant), while clusters are characterized by $kT_{200} > 2.5\,{\rm kev}$. 
Adopting different temperature profiles for clusters and groups \citep[given by eqs. 6 and 8 from][]{Yates2017}, we estimate ${T_{\rm 500, SAG}}$ for central galaxies, regardless their star formation state, with the condition that their $M_{\rm halo} \geq 10^{13}$ \msun, to make a fair comparison with observations.
Thus, the stellar mass of central galaxies within haloes of observed gas temperature ${T_{\rm 500}}$ is obtained from the connection between $T_{\rm 500, SAG}$ and \mstar~found from the model.
%

\bsp    
\label{lastpage}
\end{document}